\documentclass{jpp}

\usepackage{graphicx}
\usepackage{natbib}

\ifCUPmtlplainloaded \else
  \checkfont{eurm10}
  \iffontfound
    \IfFileExists{upmath.sty}
      {\typeout{^^JFound AMS Euler Roman fonts on the system,
                   using the 'upmath' package.^^J}%
       \usepackage{upmath}}
      {\typeout{^^JFound AMS Euler Roman fonts on the system, but you
                   dont seem to have the}%
       \typeout{'upmath' package installed. JPP.cls can take advantage
                 of these fonts, if you use 'upmath' package.^^J}%
      }
  \else
  \fi
\fi

\ifCUPmtlplainloaded \else
  \checkfont{msam10}
  \iffontfound
    \IfFileExists{amssymb.sty}
      {\typeout{^^JFound AMS Symbol fonts on the system, using the
                'amssymb' package.^^J}%
       \usepackage{amssymb}%
         
         \let\geq=\geqslant
      }{}
  \fi
\fi

\ifCUPmtlplainloaded \else
  \IfFileExists{amsbsy.sty}
    {\typeout{^^JFound the 'amsbsy' package on the system, using it.^^J}%
     \usepackage{amsbsy}}
    {}
\fi

\newcommand{\hydra}{HYDRA}
\newcommand{\zuma}{ZUMA}
\newcommand\Zeff{\ensuremath{Z_\mathrm{eff}}}
\newcommand\gcc{\nobreak\mbox{$\;$g\,cm$^{-3}$}}
\newcommand\mgcc{\nobreak\mbox{$\;$mg\,cm$^{-3}$}}
\newcommand\ep{\epsilon}

\newsavebox{\astrutbox}
\sbox{\astrutbox}{\rule[-5pt]{0pt}{20pt}}

\newcommand\p{\ensuremath{\partial}}

\title[Imposed Magnetic Field and Hot Electrons]{Imposed magnetic field and hot electron propagation in inertial fusion hohlraums}

\author[D. J. Strozzi]%
{D\ls A\ls V\ls I\ls D\ls \ns J.\ns S\ls T\ls R\ls O\ls Z\ls Z\ls I\ls$^1$%
  \thanks{Email address for correspondence: strozzi2@llnl.gov},\ns
L.\ns J.\ns P\ls E\ls R\ls K\ls I\ls N\ls S$^1$,
M.\ns M.\ns M\ls A\ls R\ls I\ls N\ls A\ls K$^1$,
D.\ns J.\ns L\ls A\ls R\ls S\ls O\ls N$^1$,
J.\ns M.\ns K\ls O\ls N\ls I\ls N\ls G$^1$,
and B.\ns G.\ns L\ls O\ls G\ls A\ls N$^1$}

\affiliation{$^1$Lawrence Livermore National Laboratory, Livermore, CA 94550, USA\\[\affilskip]}

\pubyear{2015}
\volume{650}
\pagerange{119--126}
\date{?; revised ?; accepted ?. - To be entered by editorial office}
\begin{document}

\maketitle
\begin{abstract}
  The effects of an imposed, axial magnetic field
$B_{z0}$ on hydrodynamics and energetic electrons in inertial confinement fusion (ICF) indirect-drive
hohlraums are studied.  We present simulations from the radiation-hydrodynamics code \hydra\
  of a low-adiabat ignition design for the National Ignition Facility
  (NIF), with and without $B_{z0}=70$ Tesla.
  The field's main hydrodynamic effect is to significantly reduce electron thermal conduction perpendicular to the field.  This results in hotter and less dense plasma
  on the equator between the capsule and hohlraum wall.  The inner laser beams
  experience less inverse bremsstrahlung absorption before reaching
  the wall. The x-ray drive is thus stronger from the equator with the
  imposed field.  We study superthermal, or ``hot,'' electron
  dynamics with the particle-in-cell code \zuma, using plasma conditions
  from \hydra.  During the early-time laser picket, hot electrons
  based on two-plasmon decay in the laser entrance hole
  \citep{regan-hote-pop-2010} are guided to the capsule by a 70 T
  field.  12x more energy deposits in the deuterium-tritium (DT) fuel.
  For plasma conditions early in peak laser power, we present
  mono-energetic test-case studies with \zuma\, as well as sources
  based on inner-beam stimulated Raman scattering.  The effect of the
  field on DT deposition depends strongly on the source location,
  namely whether hot electrons are generated on field lines that
  connect to the capsule.
\end{abstract}


\section{Introduction}
\label{sec:intro}

\begin{figure}
  \centerline{\includegraphics[height=5cm]{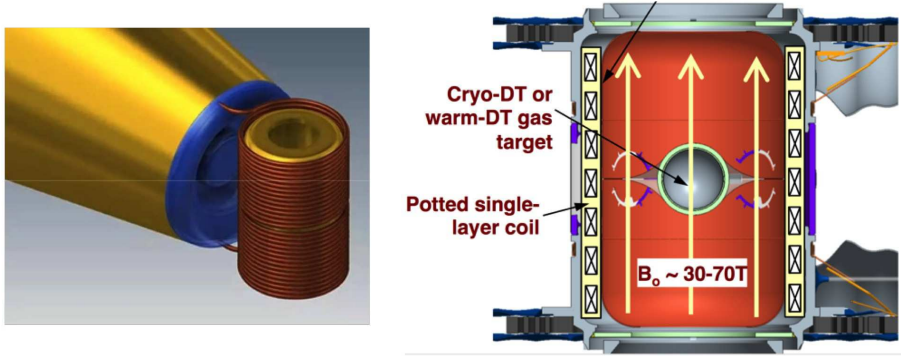}}
  \caption{Sketch of pulsed-power coil design to impose axial magnetic field on NIF hohlraum. Left: hohlraum surrounded by solenoidal coil, fielded on a Diagnostic Instrument Manipulator (DIM).  Additional hardware needed for fielding has been removed.  Right: diagram of hohlraum and solenoid, with full fielding hardware included.  The red region indicates the gold or other high-$Z$ hohlraum wall, while the gray regions outside the coil are additional support structure.}
\label{fig:hohlcoil}
\end{figure}

Using a magnetic field to enhance inertial fusion is an old idea
\citep{jones-mead-magicf-nf-1986} receiving renewed interest
\citep{slutz-maglif-prl-2012}. An imposed field is being investigated
at LLNL as a way to improve capsule performance and achieve ignition
on NIF \citep{perkins-magicf-pop-2013, perkins-maghohl-dpp-2014,
  ho-privcom-capmhd-2015}.  These simulation studies show an initial
field of 40-70 T increases both the likelihood of ignition and the
fusion yield by reducing electron-heat and alpha-particle loss from
the hot spot. Earlier experiments at the Omega laser facility with an
imposed 8 T axial field show increased fusion yield and ion
temperature in spherical implosions \citep{chang-sphere-prl-2011,
  hohenberger-b-omega-pop-2012}. The field may also limit hydrodynamic
(e.g. Rayleigh-Taylor) instability growth, and reduce the negative
effects of the growth that does occur.  The field also increases the
plasma temperature in the underdense hohlraum fill, which could reduce
SRS and improve laser propagation to the wall
\citep{montgomery-magnetized-pop-2015}. A pulsed-power approach is
being developed to impose $B_{z0}= 70$ T on a NIF hohlraum
\citep{rhodes-magnifico-ieee-2015}, and is sketched in Fig.\
\ref{fig:hohlcoil}. Laser-driven capacitor-coil systems are a possible
way to impose 100-1000 T fields \citep{fujioka-B-scirep-2013,
  pollock-bfield-dpp-2014}.

This paper presents simulation studies of how an imposed field affects
hohlraum hydrodynamics and energetic electrons. First, we report on
simulations using the radiation-hydrodynamics code \hydra\
\citep{marinak-hydra-pop-2001} with and without an imposed field of
ignition experiment N120321.
Then we show studies with the particle-in-cell code \zuma\
\citep{larson-zuma-dpp-2010, strozzi-fastig-pop-2012} of the field's
effect on energetic or ``hot'' electrons. 

We study NIF shot N120321, which used a 4-shock, low-adiabat or
``low-foot'' laser pulse, a plastic ablator, and a cryogenic DT ice
layer.  It achieved the highest fuel areal density to date on NIF, and
has been extensively modeled to understand its low neutron yield
\citep{clark-120321-pop-2015}. Here, we use \hydra's
magnetohydrodynamics (MHD) package \citep{koning-mhd-dpp-2006} with
$B_{z0}=70$ T, which was not present in the actual experiment.  We
include the $\vec J\times\vec B$ magnetic pressure force, a simple
Ohm's law $\vec E=\eta\vec J-\vec v\times\vec B$, Ohmic heating, and
anisotropic electron thermal conductivities parallel and perpendicular
to $\vec B$ (but not the Righi-Leduc heat flow along
$\vec B \times \nabla T_e$).  This neglects several effects which
could be important, and will be studied in future work, namely the
self-generated or ``Biermann battery''
$\p_t\vec B \propto \nabla T_e \times \nabla n_e$ field and the Nernst
effect ($\vec E \propto \vec B \times \nabla T_e$).  In our runs, the
$B$ field roughly follows the MHD ``frozen-in law'' for the
highly-conducting plasma flow.  The primary effect of the field is to
reduce electron heat conduction perpendicular to $\vec B$. This
leads to a hotter hohlraum fill, and a wider channel between the
capsule and hohlraum equator. The inner cone of beams (pointed toward
the equator) better propagate to the wall, which gives more equatorial
x-ray drive and a less oblate imploded capsule. This would reduce the
need for energy transfer to the inners, and probably reduce their
backscatter - both due to the lower power and higher temperature.

Besides hydrodynamics, we also study hot electron dynamics. Hot
electrons are a generic aspect of intense laser-plasma interactions
(LPI).  They are produced in any parametric process that drives a
Langmuir wave.  Of particular interest in ICF are stimulated Raman
scattering (SRS) and two-plasmon decay (TPD).  These are the decay of
a light wave to a Langmuir wave and, respectively, a scattered light
wave (SRS) or another Langmuir wave (TPD).  In many laser-produced
plasmas, the daughter Langmuir waves are damped primarily by
collisionless Landau damping, which entails the resonant interaction
of the wave with electrons at its phase velocity.  This is typically
greater than the electron thermal speed, and therefore produces a
population of superthermal or ``hot'' electrons.  Experiments show the
resulting hot-electron spectrum from a single parametric process is
roughly exponential with ``temperature'' $T_h$,
$dN/dE \propto g(E) e^{-E/T_h}$ ($E$ is the hot electron kinetic
energy), with $g=E^{1/2}$ for a non-relativistic Maxwellian.  NIF
experiments with gas-filled hohlraums have shown hard x-ray output
consistent with a two-temperature hot-electron population, attributed
to Raman backscatter, and TPD or SRS at quarter-critical
density \citep{doppner-hote-prl-2012}.  Relativistic processes that
produce $>$ MeV electrons at intensities $I\lambda^2 > 10^{18}$ W
cm$^{-2}$ $\mu$m$^2$ are of great interest in the short-pulse and
fast-ignition fields, but are not discussed here.

This paper focuses on hot electrons in ignition hohlraums, though
similar considerations apply to directly-driven targets.  Hot
electrons impede ICF in several ways, namely implosion asymmetry and
fuel preheat. The laser power transferred to hot electrons generally
stays in the target, so is not a power loss like backscattered light.
But, the deposition in space and time differs from the intended
inverse-bremsstrahlung absorption of the incident laser.  NIF
hohlraums with high hohlraum gas fill density ($\gtrsim 0.9$ \mgcc\
He) have generally shown large SRS from the inner beams.  This reduces
the inner-beam power reaching the wall -- both by scattered light and
Langmuir waves -- which makes the implosion more oblate (or
``pancaked'').  The Langmuir wave energy remains in the target, but
heats the hohlraum wall by conduction in a much larger area than the
inner-beam spots.  To control symmetry, cross-beam energy transfer has
been used to move power from the outer to inner beams, inside the
target \citep{michel-xbeam-prl-2009}. Hot electrons with energy
$\gtrsim 170$ keV can also preheat the fusion fuel (e.g., cryogenic DT
ice layer) by depositing energy separate from the intended shock
sequence and capsule compression \citep{salmonson-hots-dpp-2010,
  haan-nic-pop-2011}.  This results in a higher fuel adiabat, which
significantly reduces the achievable compression.

We propagate hot electrons with the hybrid-PIC code \zuma\ through
plasma conditions from \hydra.  We run \zuma\ in a ``Monte-Carlo
mode'' with no $E$ or $B$ fields, except sometimes a static $B$.  Hot
electrons undergo energy loss and angular scattering as they
propagate, and the energy deposition profile is found with
and without an initial $B_{z0}$. We first present an unphysical
test-case study of mono-energetic hot electrons directly incident on
the capsule (unrealistic for LPI-produced hot electrons) early in peak
laser power (time 18 ns). A minimum initial energy $E_0=$125 keV is
needed to penetrate the ablator and reach the DT layer.  The maximum
energy deposited in the DT layer, $E_{DT}$, occurs for $E_0=$185 keV
and is $E_{DT}/E_0=13\%$.  Higher energy electrons do not fully stop
in DT.

We then examine a realistic hot-electron source, consistent with
two-plasmon decay during the early-time ``picket'' or initial part of
the laser pulse (time 1 ns).  The deposition is mostly in the high-$Z$
wall, as expected from solid-angle arguments, and
$E_{DT}/E_h = 2.2\times 10^{-3}$ with $E_h$ the total injected hot
electron energy.  Adding a uniform 70 T axial $B$ field strongly
magnetizes the hot electrons in the hohlraum fill, guides them to the
capsule, and increases $E_{DT}/E_h$ by 12x to 0.026. This may not
degrade fusion performance, since NIF experiments have shown greatly reduced
picket hot electrons with pulse shaping, e.g.\ a low-power ``toe'' to
burn down the window \citep{dewald-hots-prl-2015, moody-aps-pop-2014}.

Finally, we consider a hot-electron source consistent with SRS of the
inner laser beams, early in peak laser power (18 ns). With no
$B_{z0}$, the hot electrons deposit throughout the target, with a very
small $E_{DT}/E_h\approx 1.2\times 10^{-4}$. With $B_{z0}=70$ T, the
field strongly magnetizes the hot electrons in the hohlraum fill gas.
The deposition in DT is greatly increased (decreased) for hot
electrons originating on field lines that do (do not) connect to the
capsule at this time.

This paper is organized as follows.  We describe our MHD simulation
methodology in section \ref{sec:mhdmethod} and our MHD results in
section \ref{sec:mhdres}. The \zuma\ simulation method is detailed in
section \ref{sec:zumamethod}. Section \ref{sec:captest} discusses test
cases of mono-energetic electron propagation through the capsule.
Section \ref{sec:picket} presents \zuma\ results for a TPD-relevant
source during the picket, and shows a 12x increase in $E_{DT}$ with a
70 T axial field.  In section \ref{sec:srspeak} we present \zuma\
results early in peak laser power with an SRS-relevant hot electron
source, using plasma conditions and the $B$ field from our \hydra\
simulations, and find a strong dependence in $E_{DT}$ on source location.  We
conclude in section \ref{sec:conc}.

\section{\hydra\ MHD simulation method}
\label{sec:mhdmethod}

\begin{figure}
  \centerline{\includegraphics[height=6cm]{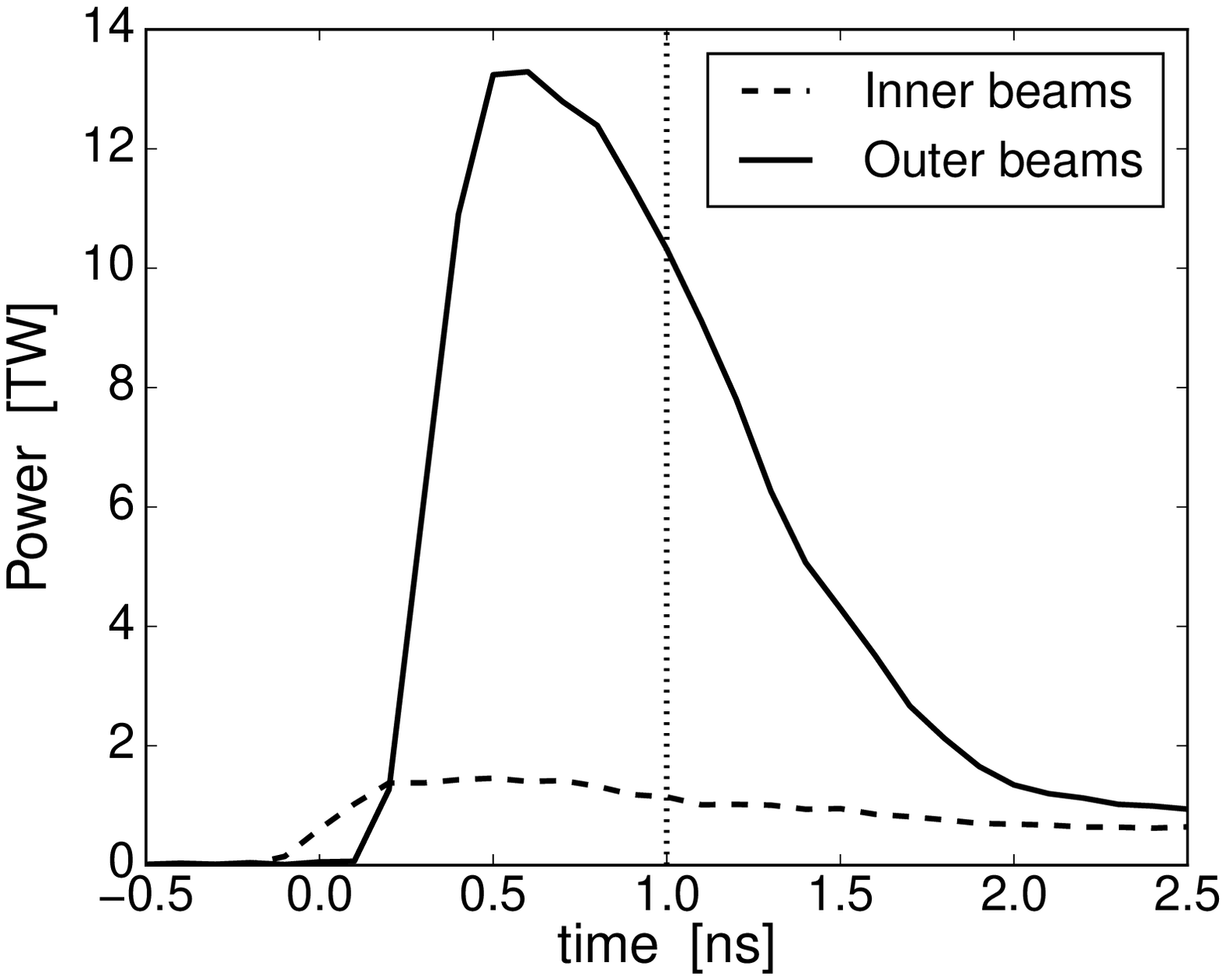}
    \includegraphics[height=6cm]{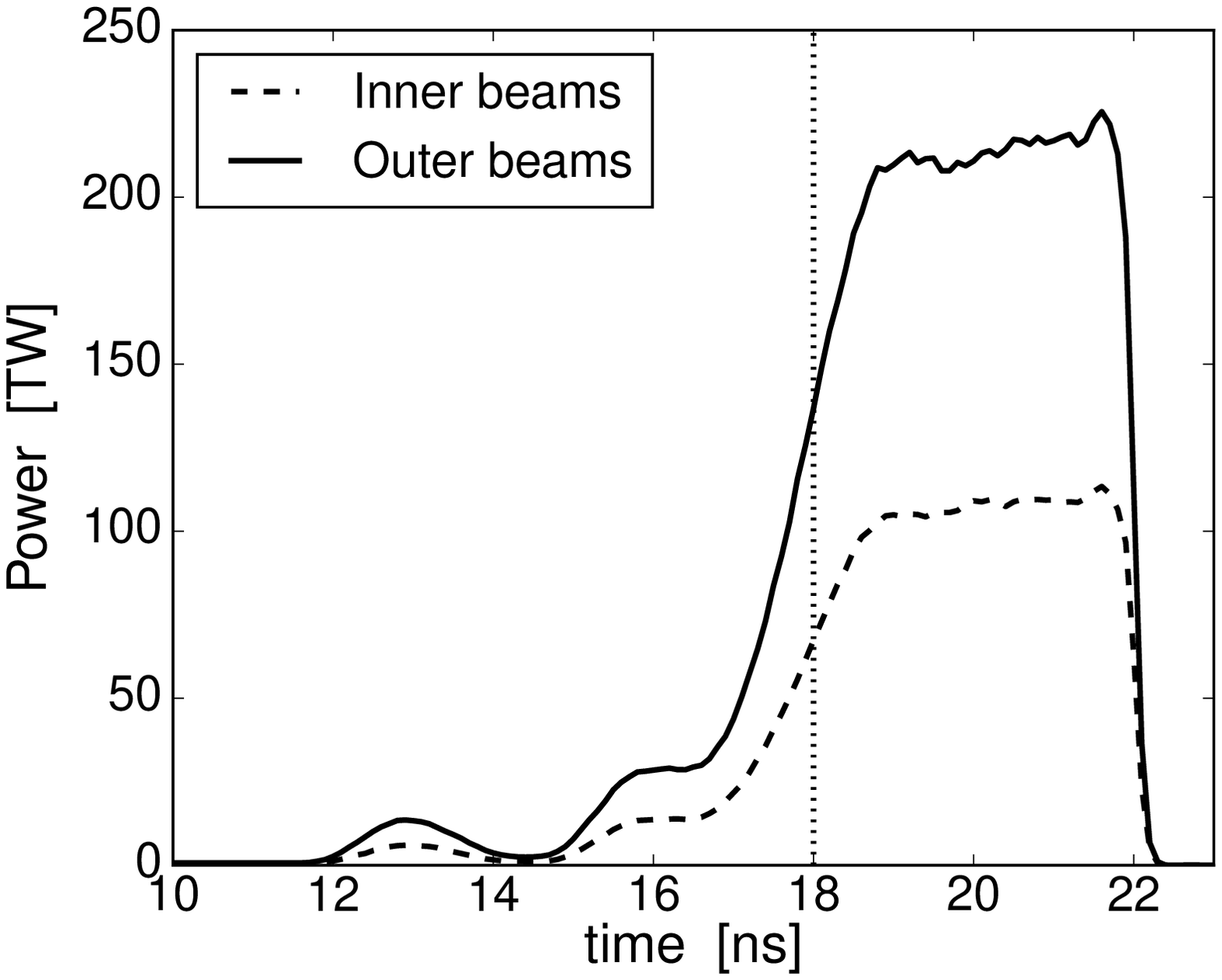}}
  \caption{Incident laser power on NIF shot N120321 on the inner (dashed) and
    outer (solid) laser beams during the early-time picket (left) and peak power
    (right). The dotted vertical lines at 1 and 18 ns indicate
    times \hydra\ plasma conditions are used for \zuma\ hot electron studies.}
\label{fig:pulse}
\end{figure}

We use the radiation-hydrodynamics code \hydra\ to simulate the NIF
hohlraum experiment N120321.  This shot used a 4-shock, ``low-foot''
laser pulse (shown in Fig.\ \ref{fig:pulse}), a plastic ablator
(C$_{0.42}$H$_{0.57}$ plus small amounts of O impurity, and Si dopant
to control x-ray preheat) with a DT ice layer, and a depleted uranium
(DU) hohlraum with a thin 0.7 $\mu$m inner gold coating.  The hohlraum
fill gas was 0.96 \mgcc of He. The methodology is the standard one in
use for hohlraum simulations at LLNL \citep{jones-mults-pop-2011},
entailing the ``high-flux model'' with detailed configuration
accounting (DCA) for non-LTE material properties, and an electron
thermal flux limit of 0.15 the free-streaming value
\citep{rosen-dca-hedp-2011}.  The runs use a 3D mesh with one zone and
periodic boundary conditions in azimuth, and are effectively
cylindrical 2D (r-z).  The mesh is managed with an arbitrary
Lagrangian-Eulerian (ALE) approach, designed to keep the mesh as
Lagrangian as possible. We use the full, incident laser energy of 1.52
MJ, and neither remove measured backscatter nor degrade the laser
power as is needed to match x-ray drive data in high gas fill targets
\citep{jones-mults-pop-2011}.  Our \hydra\ simulations are therefore
not proper post-shots, but address the role of an imposed field in
ignition-scale designs, and provide relevant plasma conditions for hot
electron studies.

A distinct aspect of the present work is the inclusion of an initial
axial magnetic field $B_{z0}$.  \hydra's magnetohydrodynamics (MHD)
package \citep{koning-mhd-dpp-2006} was used to model $B_z$ and the
resulting $B_r$ required by $\nabla\cdot\vec B=0$.  No azimuthal field
is produced for our axisymmetric geometry and simple Ohm's law. The MHD package uses a 3D finite element method, with appropriate boundary conditions to be effectively axisymmetric. We used the Ohm's law 
\begin{equation}
\vec E = \eta\vec J-\vec v\times\vec B
\end{equation} 
where $\eta$, $\vec J$, and $\vec v$ are scalar resistivity, net current, and center-of-mass velocity, respectively.  The $B$ field is evolved via $\partial_t\vec B=-\nabla\times\vec E$. The MHD package as used here affects the matter in three ways: a) the $\vec J\times\vec B$ force, b) $\eta J^2$ (Ohmic) and other heating terms, and c) a tensor electron thermal conductivity:
\begin{equation}
  \mathsfbi\kappa = \kappa_\perp\left(\mathsfbi I-\hat b\hat b \right) + \kappa_{||} \hat b\hat b \qquad \hat b \equiv {\vec B \over |\vec B|}.
\end{equation}
The Righi-Leduc effect, with a separate conductivity along $\vec B\times\nabla T_e$, is currently neglected($T_e$ is electron temperature).  An artificial flux limit $f$ is imposed, as is typical in hohlraum simulations.  Specifically, the component of the heat flux $\vec q_e $ along each logical index (not physical) coordinate $\hat i$ is limited: $\vec q_e\cdot\hat i = \mathrm{min} [ \hat i \cdot \mathsfbi\kappa\cdot \nabla T_e, fq_{FS}]$ where $q_{FS} \equiv n_eT_e^{3/2}/m_e^{1/2}$ is the free-streaming heat flux.

The anisotropic heat conduction has the largest effect in our simulations. We expect $B\gtrsim 1$ T to significantly reduce $\kappa_\perp$ below its unmagnetized value $\kappa_{||}$. The Hall parameter $H\equiv\omega_{ce}\tau_{ei}$ for thermal electrons is
\begin{subeqnarray}
  H &\equiv& \omega_{ce}\tau_{ei} = {B \over B_0}, \\
 B_0 &\equiv& {(32\pi)^{1/2} \over 3}{m_e^{1/2}e^3 \over (4\pi\epsilon_0)^2} {n_e \over T_e^{3/2}} \Zeff, \\
\Zeff &\equiv& {\sum_if_iZ_i^2\ln\Lambda_{ei} \over \sum_if_iZ_i}. \label{eq:zeff}
\end{subeqnarray}
For each ion species $Z_i$ is the ionic (not  nuclear) charge, $n_i=f_in_I$,
and $n_I=\sum_in_i$ is the total ion number density.  In practical units, $B_0[\mathrm{T}] = 4.73 (n_e/n_{cr})\Zeff/T_e[\mathrm{keV}]^{3/2}$ with $n_{cr}=9.05\times 10^{21}$ cm$^{-3}$ the critical density for light of wavelength 351 nm.  For $Z_i=2$ He at $n_e=0.1n_{cr}$ and $T_e=3$ keV, typical of the underdense hohlraum fill, we find $\ln\Lambda_{ei}=7.9$ and $B_0=1.43$ T. Given $B_{z0}=70$ T, most of the underdense plasma fill should be strongly magnetized. $\kappa_\perp$ decreases with $H$ according to
\begin{equation}
  {\kappa_\perp \over \kappa_{||}} \approx {1+p_1H \over 1+p_2H+p_3H^2+p_4H^3},
\end{equation}
with the $Z_i$-dependent fitting coefficients $p_j$ given in \citet{epperlein-xport-pof-1986}. For $Z_i=2$ He, $\kappa_\perp/\kappa_{||}=0.1$ for $H=1.6$. $\kappa_{||}$ is found either from the \citet{lee-more-pof-1984} formulation to include dense-plasma effects, or interpolation from an advanced table. The Epperlein and Haines results are used to include electron self-collisions ($Z_i<\infty$) and dependence on $H$.

\section{MHD simulation results}
\label{sec:mhdres}

\begin{figure}
  \centerline{\includegraphics[height=6cm]{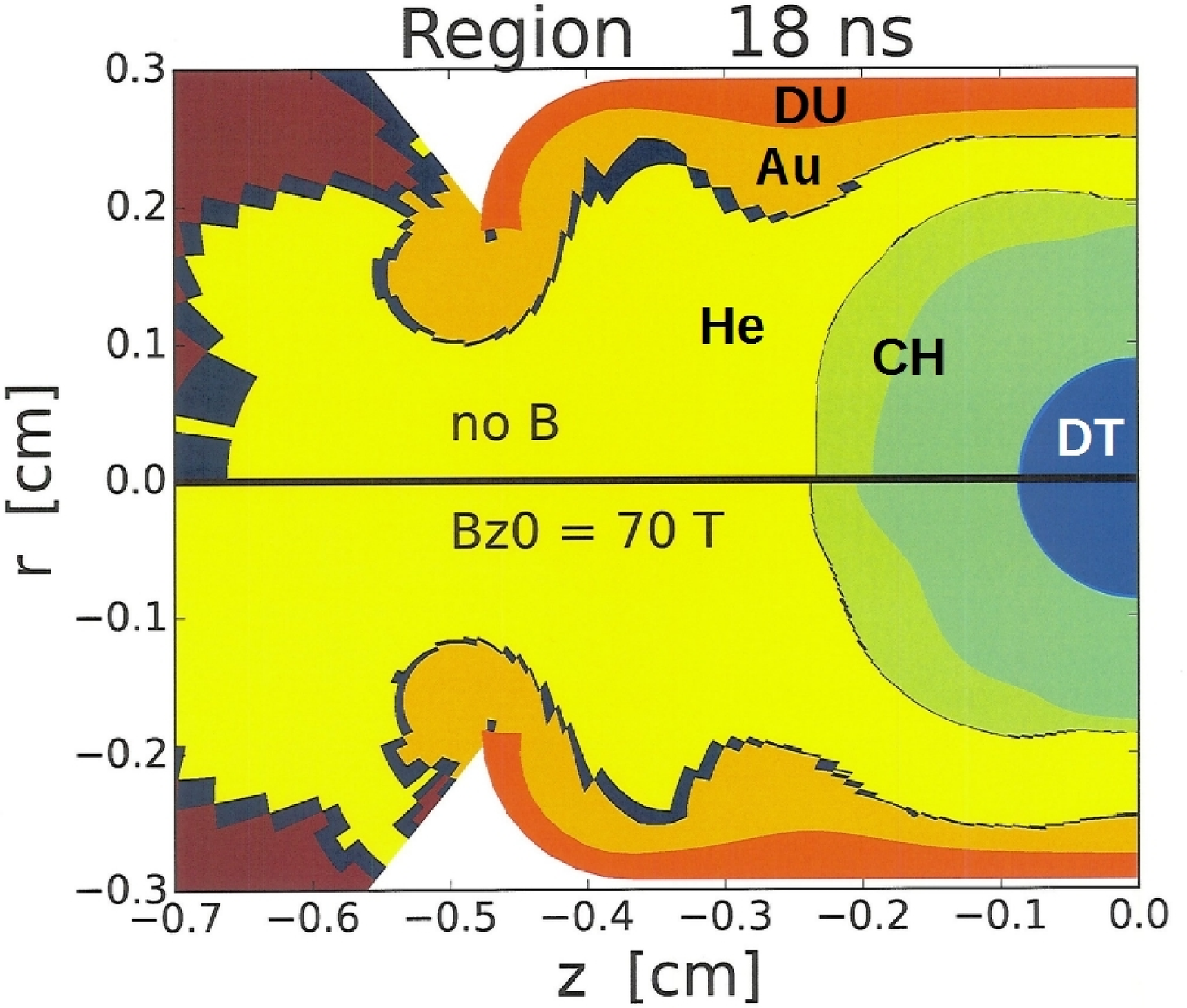}
  \includegraphics[height=6cm]{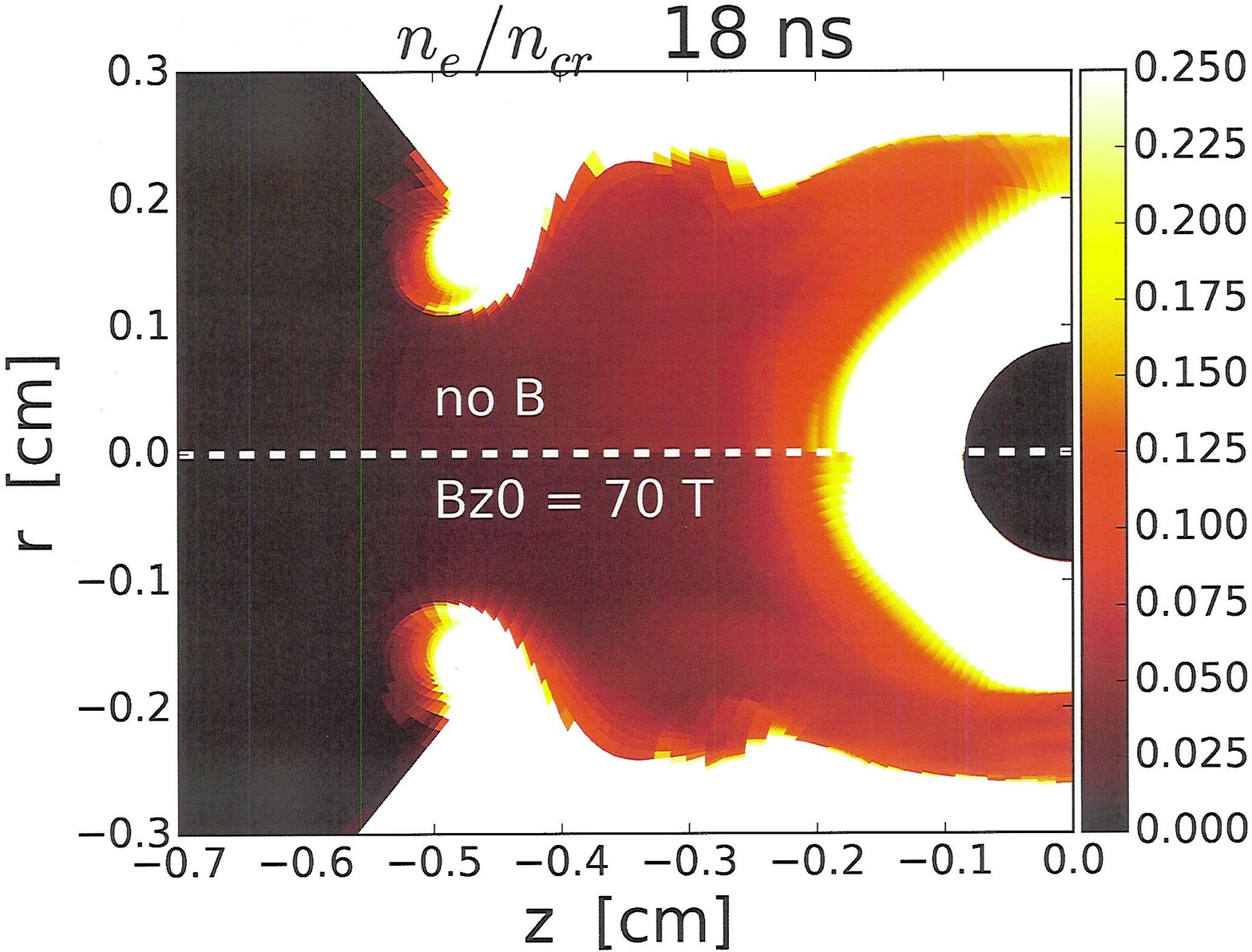}}
  \caption{Plasma conditions from \hydra\ simulations of NIF shot N120321 at 18 ns, used in SRSPEAK \zuma\ run series. The left $(z<0)$ half of the hohlraum is plotted, but the simulation included both halves.  Top half ($r>0$): without MHD, bottom half ($r<0$): with MHD and initial axial magnetic field $B_{z0}=70$ T. Left: material region: DT, He, CH (two green regions), Au, DU label deuterium-tritium, helium, plastic, gold, and depleted uranium.  Right: free electron density in units of critical density for 351 nm light.}
\label{fig:18nsmaps}
\end{figure}

\begin{figure}
  \centerline{\includegraphics[height=6cm]{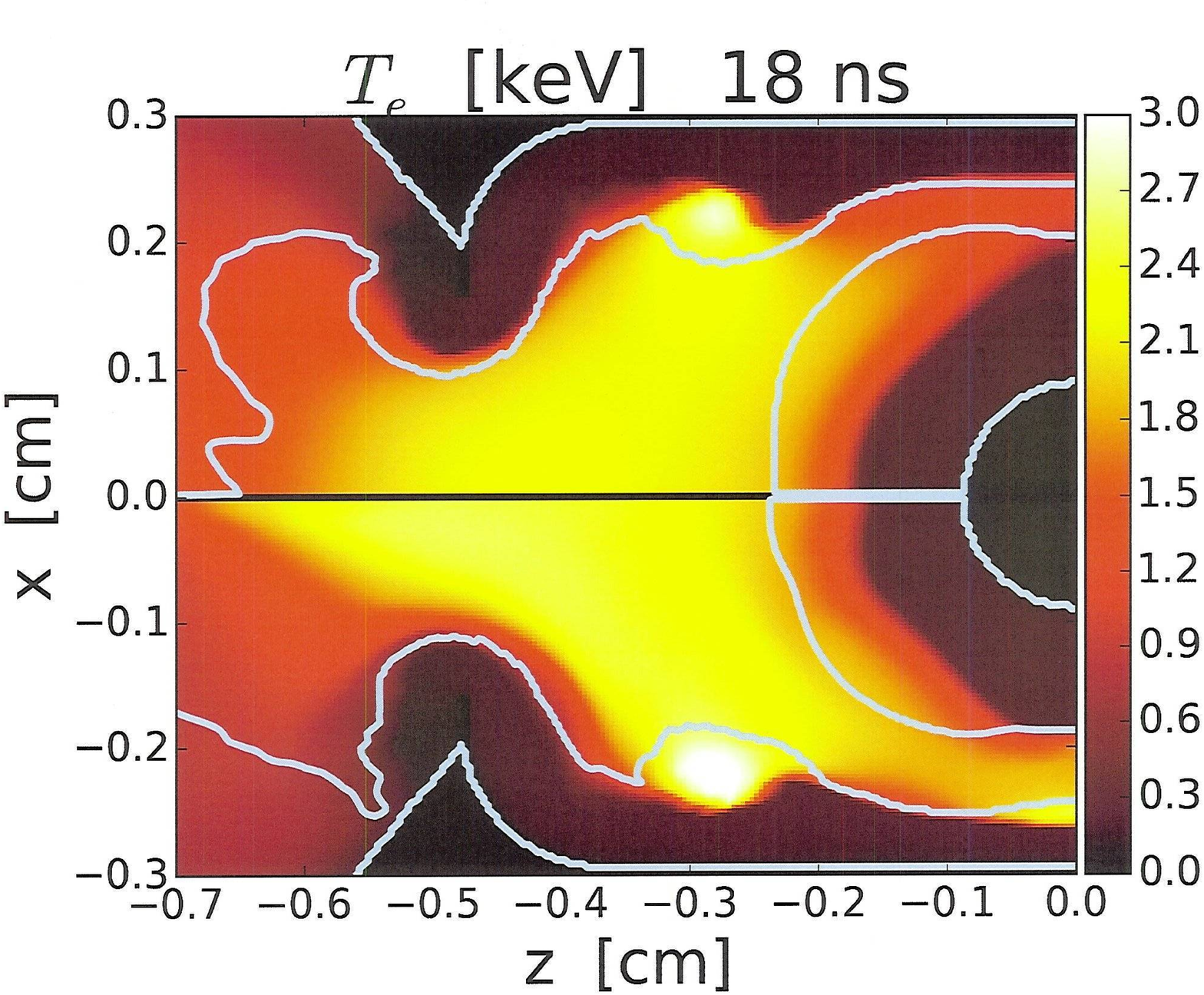}
  \includegraphics[height=6cm]{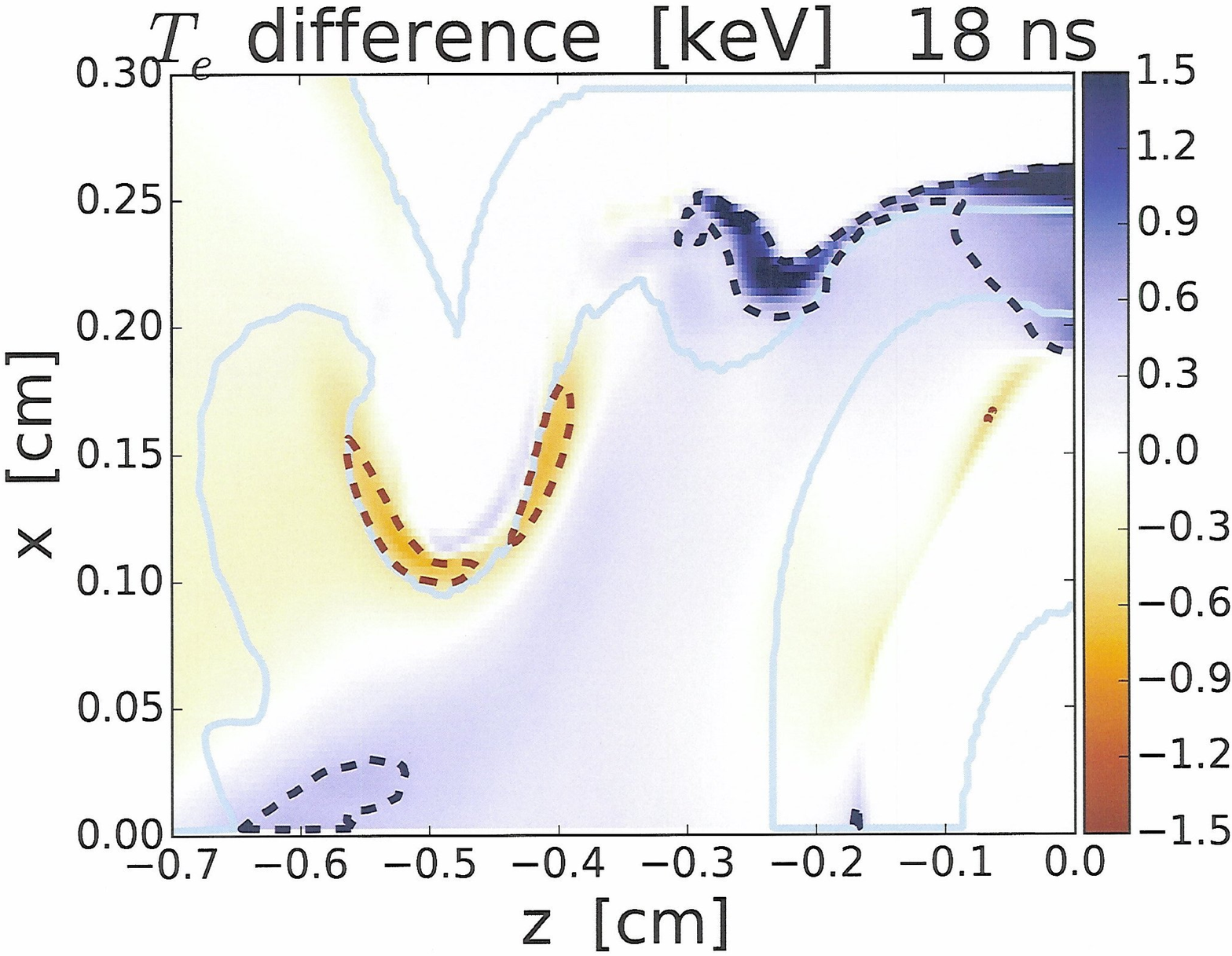}} 
  \centerline{\includegraphics[height=6cm]{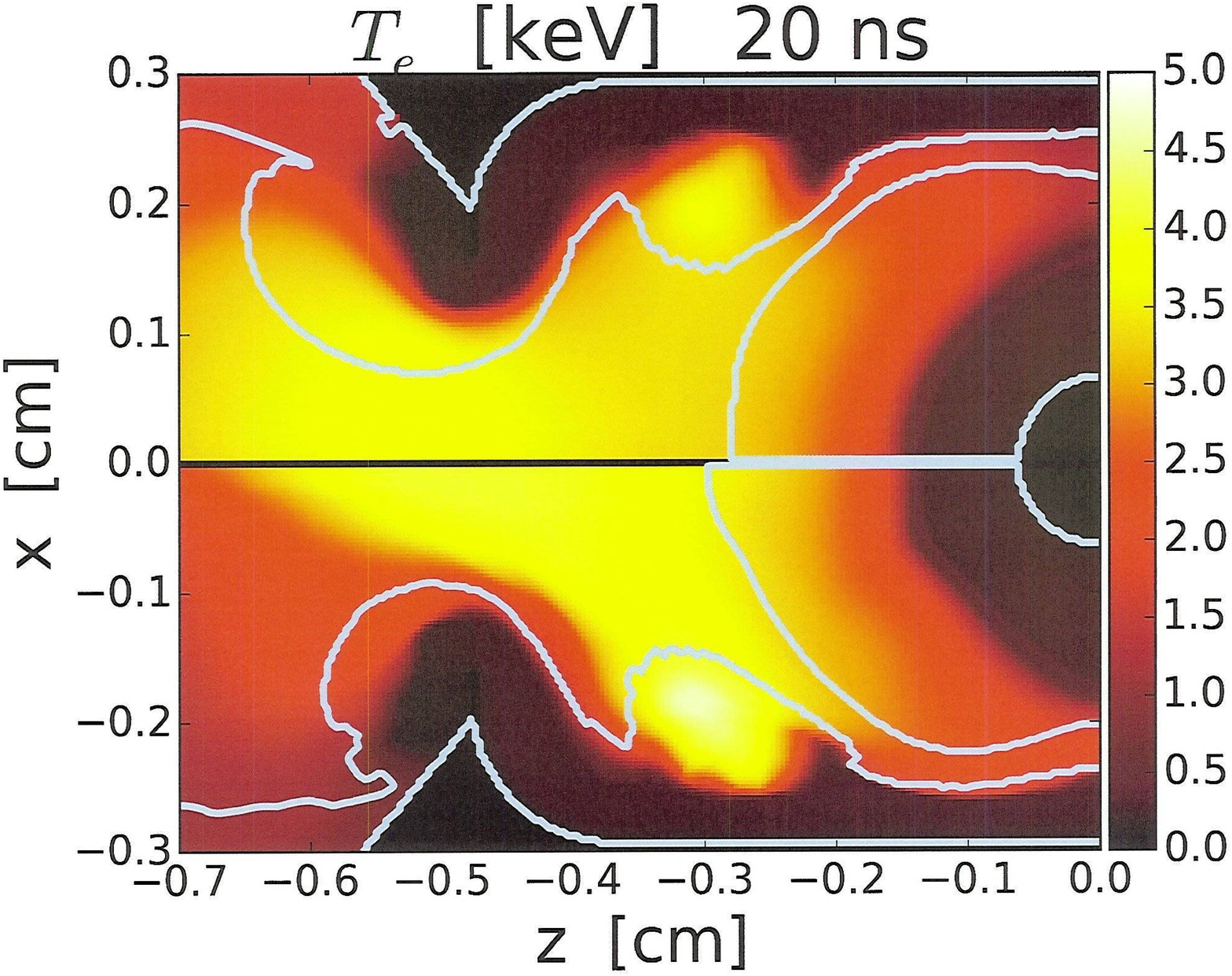}
  \includegraphics[height=6cm]{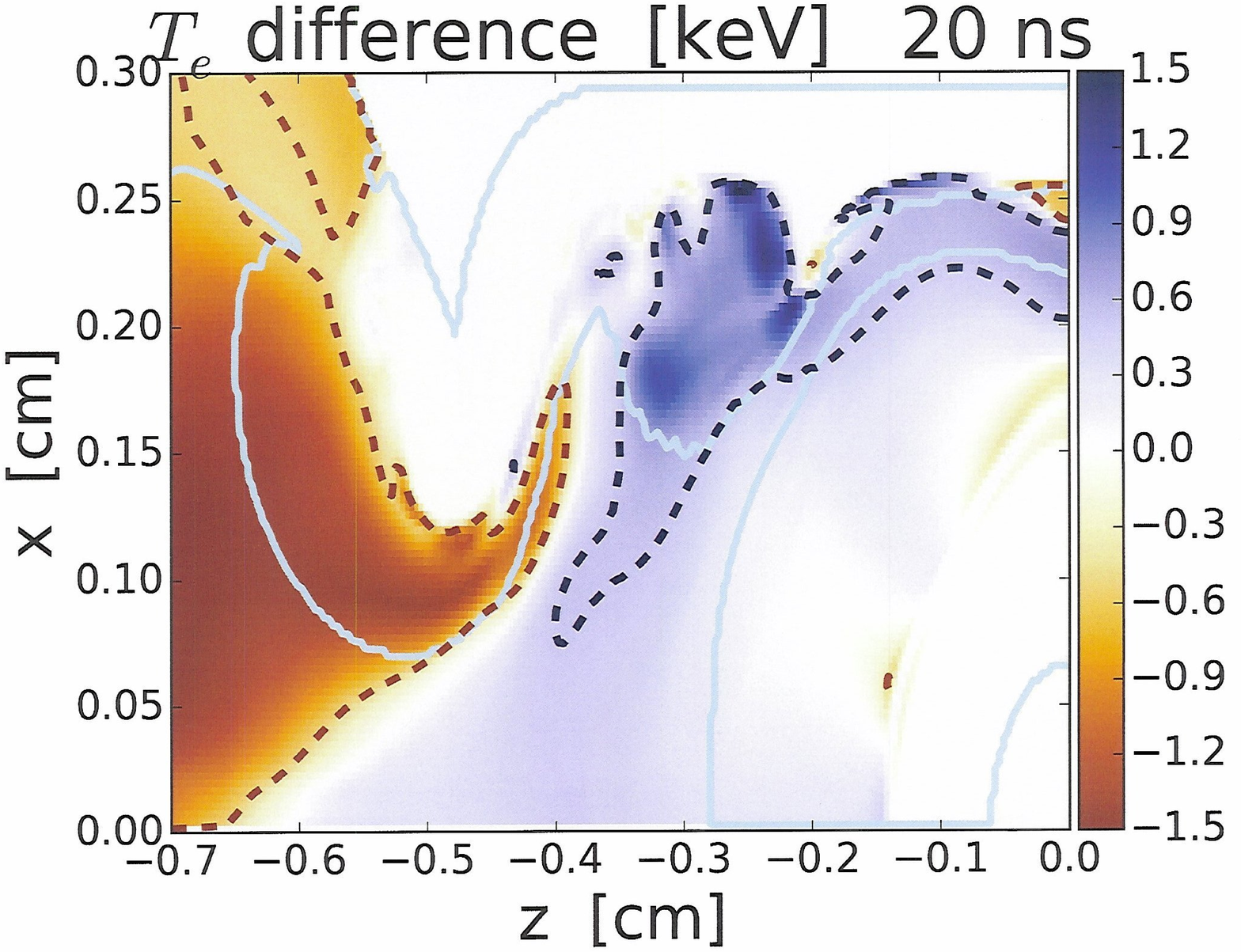}} 
  \centerline{\includegraphics[height=6cm]{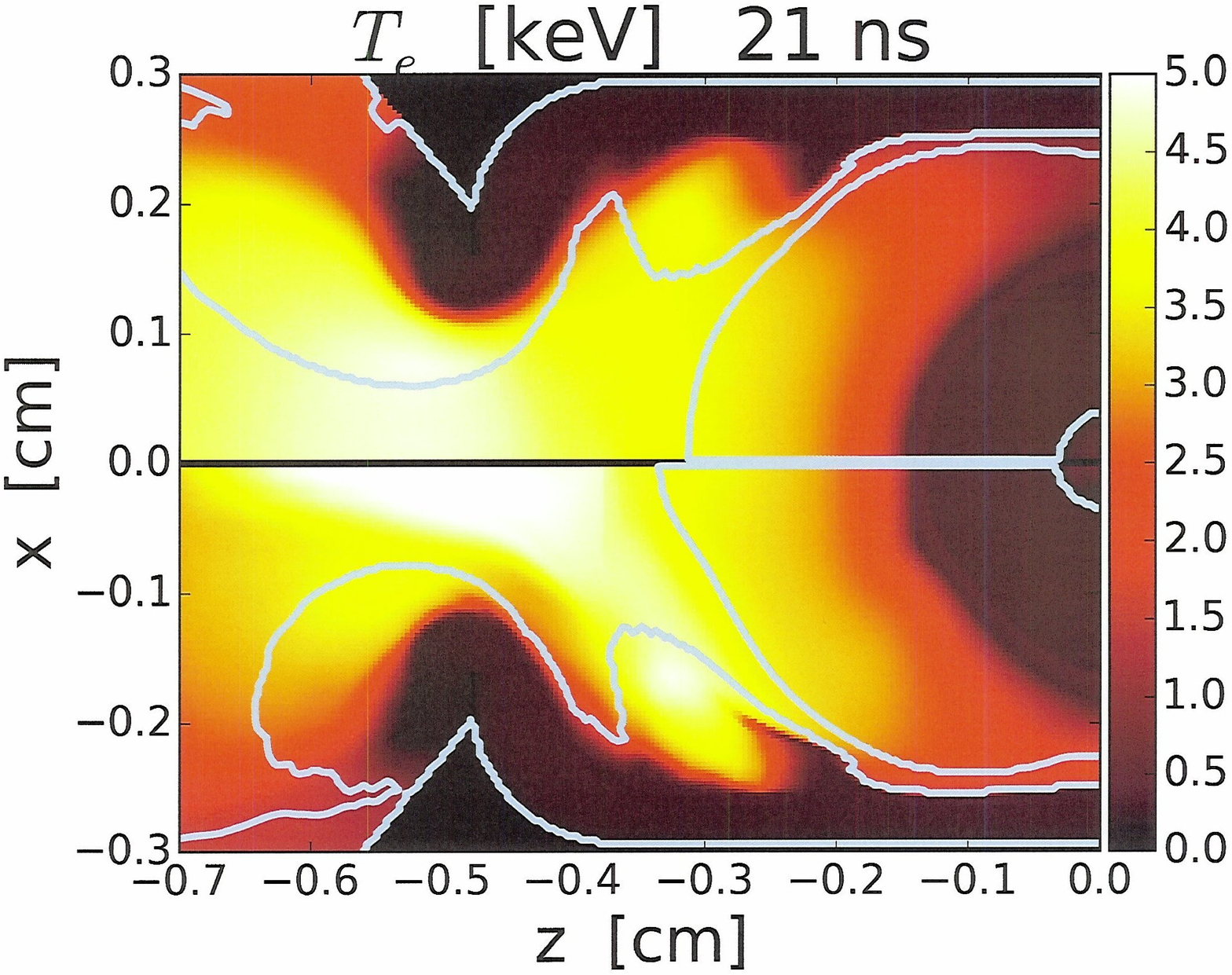}
  \includegraphics[height=6cm]{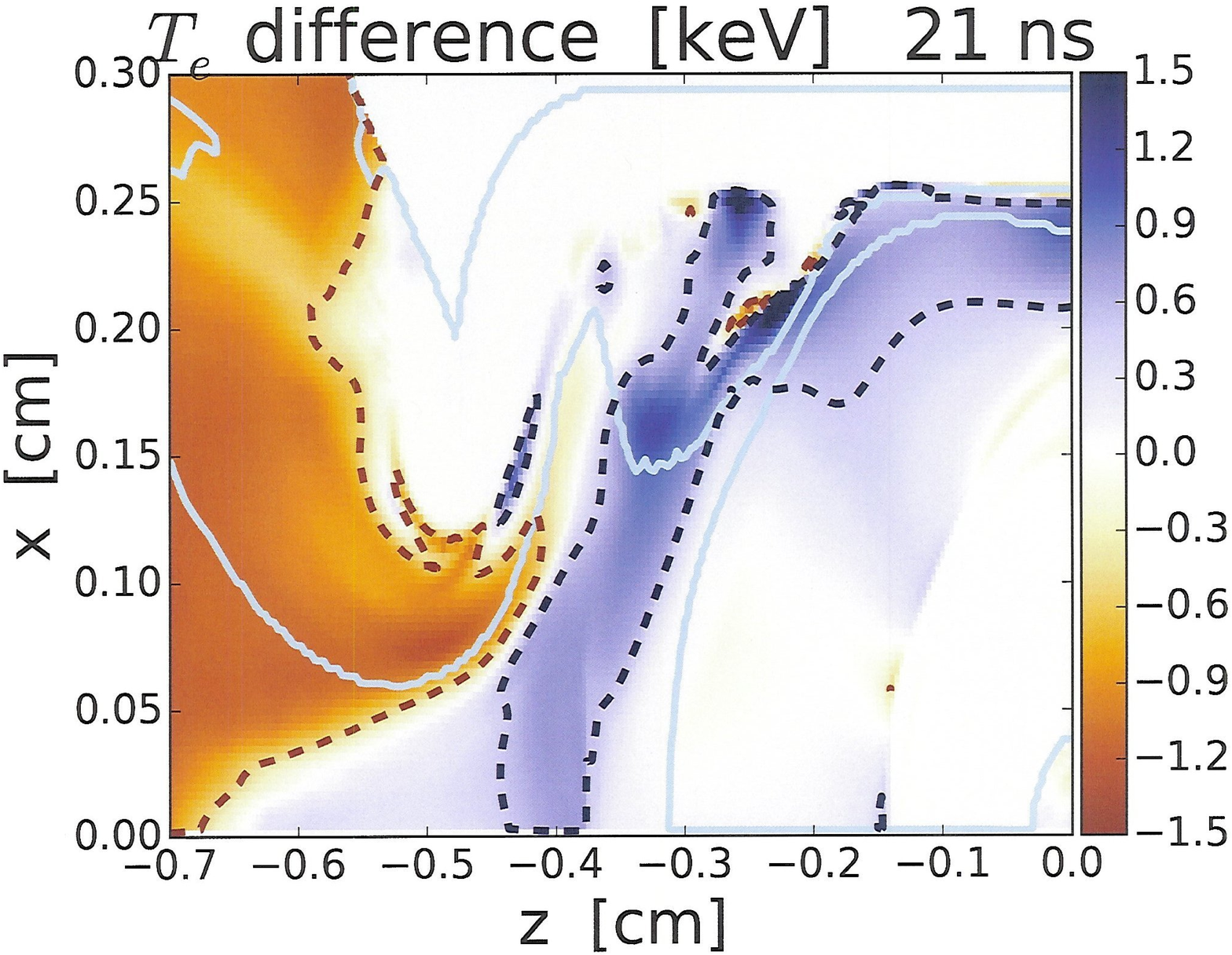}} 
\caption{Left: electron temperature in keV from \hydra\ MHD simulations shown in Fig.\ \ref{fig:18nsmaps}, at several times.  Right: temperature difference with MHD minus without MHD.  Light blue contour marks region boundary for DT, CH, and Au.  Dashed contours are temperature differences of $\pm0.5$ keV.}
\label{fig:Temaps}
\end{figure}

\begin{figure}
  \centerline{\includegraphics[height=5cm]{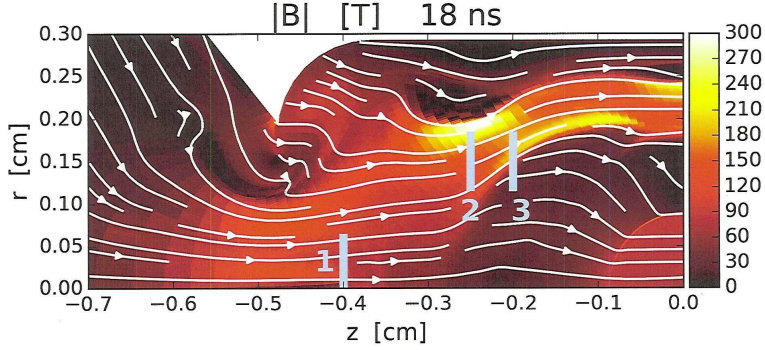}}
  \caption{$|\vec B|$ at 18 ns from \hydra\ MHD simulation shown in Fig.\ \ref{fig:18nsmaps}. The dark-red color in the capsule is the uncompressed field of 70 T. White curves are stream lines (integral curves) of the vector field $(B_z,B_r)$. Light blue boxes and text indicate \zuma\ hot electron sources for runs with and without MHD.}
\label{fig:18nsB}
\end{figure}

\begin{figure}
  \centerline{\includegraphics[height=7cm]{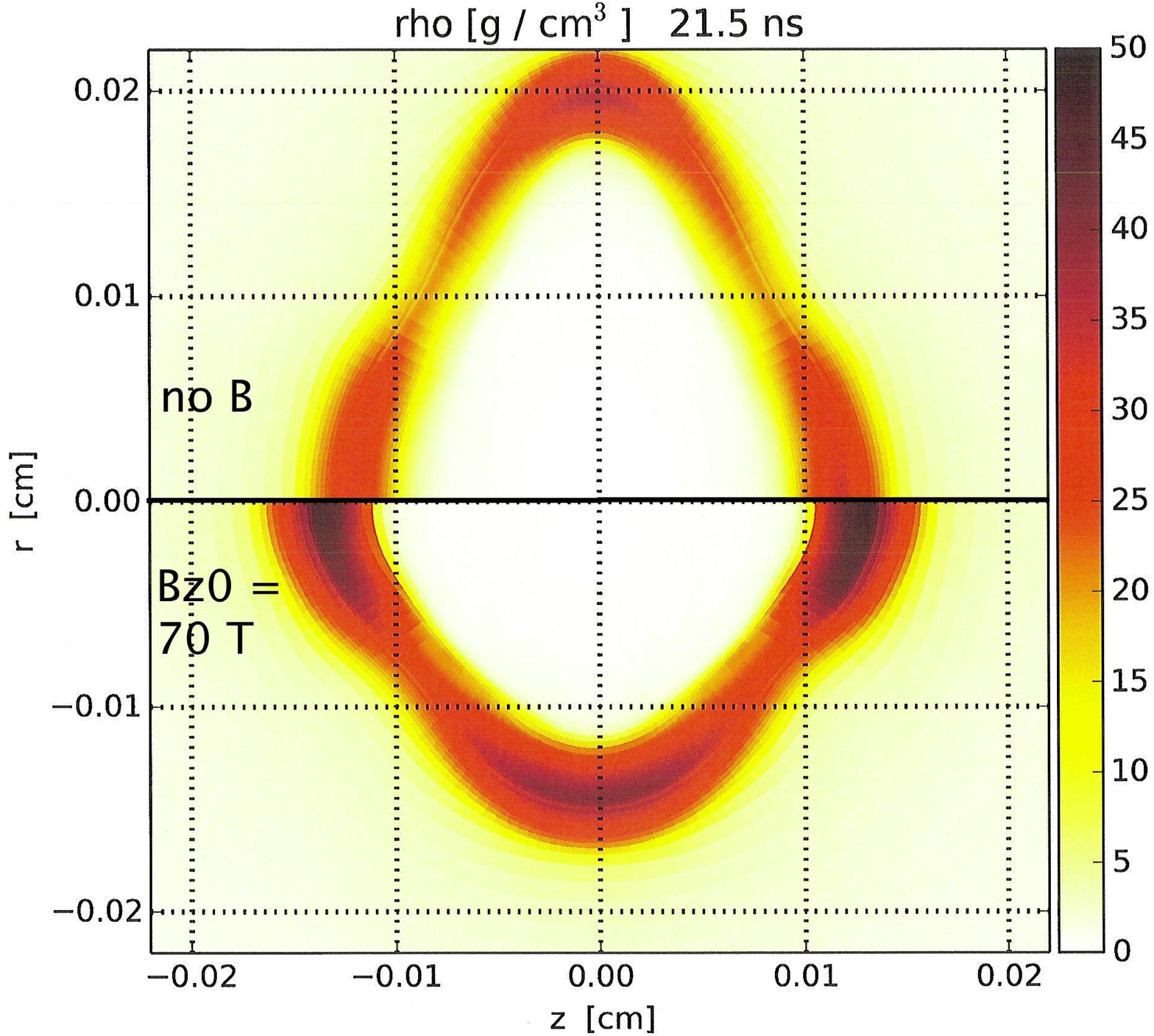}}
  \caption{Capsule shell density at 21.5 ns from \hydra\ simulations shown in Fig.\ \ref{fig:18nsmaps} without MHD (top half, $r>0$), and with MHD and $B_{z0}=70$ T (bottom half, $r<0$). The shell is oblate without MHD, while it is close to round with MHD.  This reflects the improved inner beam propagation with the field.}
\label{fig:21p5rho}
\end{figure}

The \hydra\ runs of NIF shot N120321 with and without MHD are
qualitatively similar. The principal difference is the MHD run has
higher electron temperature in some regions and a wider channel of He
fill gas at the equator. This leads to better inner beam propagation
to the wall (less inverse-bremsstrahlung absorption in the low-$Z$
fill), and results in a less oblate capsule. Figure \ref{fig:18nsmaps}
shows the material regions and electron density for the two runs at 18
ns, during the rise to peak power. We use this time for \zuma\
simulations of SRS hot electrons. A more detailed density plot is in
Fig.\ \ref{fig:18ns-necap}.  Electron temperature $T_e$ with and
without MHD at several times during peak power is displayed in Fig.\
\ref{fig:Temaps}. The field increases $T_e$ mainly in the laser-heated
gold, such as the outer-beam ``bubble'' $(r,z) \approx (0.2,-0.3)$
cm. The He gas fill is also hotter with the field, but less so than
the gold. The low density plasma outside the laser entrance hole (LEH)
is cooler with the field, though the total energy in this region is
small.

The $B$ field for the MHD run is plotted in
Fig.\ \ref{fig:18nsB}. It roughly follows the MHD frozen-in law, and
advects with the radial motion of the ablator and high-$Z$ wall. The
compressed field approaches 300 T and continues to grow with time. The
white stream line that just touches the capsule outer radius at
$(z,r)=(0,0.1)$ roughly separates field lines that are still connected
to the capsule ($r<0.1$ cm at $z=0$), from those that have advected
with the ablated blowoff and no longer connect to the capsule.

The capsule density at the end of peak power is plotted in Fig.\
\ref{fig:21p5rho}.  The dense fuel is oblate without the field, but
becomes close to round with it.

\section{\zuma\ Hot electron simulation method}
\label{sec:zumamethod}

We propagate hot electrons through fixed plasma conditions from
\hydra\ using the hybrid-PIC code \zuma\ in a ``Monte-Carlo'' mode.
We do not include forces from $E$ and $B$ fields, except when we
include a specified (static) $B$ field.  The background plasma
properties are not updated.  In other work, \hydra\ and \zuma\ have
been coupled to run in tandem, and applied to fast ignition designs
\citep{strozzi-fastig-pop-2012}. The hot electrons undergo collisional
energy loss off background electrons, and angular scattering off
background electrons and ions.  We neglect collisions among hot
electrons, since their density is much less than the background
species. We use the formulas in \citet{robinson-fet4fi-nf-2014} for a
fast electron with $v\gg v_{Te}$.  The energy loss rate (stopping
power) is given by
\begin{subeqnarray}
  \label{eq:3}
  {dE\over dt} &=& {C_e n_e \over m_e v}L_d \\
               &\approx& {C_e n_e \over \sqrt{2m_eE}} \ln{E\over\hbar\omega_p}, \qquad
               \hbar\omega_p \ll E \ll m_ec^2, \\
  L_d &=& \ln{pv \over \hbar\omega_p\sqrt{\gamma+1}} - {\ln2\over2} + {9\over16} + 
        {\ln2+1/8\over\gamma}\left({1\over2\gamma}-1\right).
\end{subeqnarray}
$C_e\equiv e^4/4\pi\ep_0^2$, the fast electron kinetic energy $E=m_ec^2\epsilon$, $\gamma=\epsilon+1$ is the Lorentz factor, and $p=\gamma m_ev$. $n_e$ is the total (free plus bound) background electron
density. $L_d$ given above is valid for energy loss off free electrons, or bound electrons
for sufficiently high $E$ or $n_e$ (the ``density effect''). This assumption may not be valid for all electrons.  The angular scattering rate is
\begin{subeqnarray}
  \label{eq:dta2dt}
{d\left\langle\theta^2\right\rangle \over dt} &=& {2C_e \over p^2v} 
    \left[ n_I \left\langle Z^2 \right\rangle L_{si} + n_{e,f}L_{se} \right]
  \\
 &\approx&  {C_e \over \sqrt{2m_e} E^{3/2}} \left[ n_I \left\langle Z^2 \right\rangle + n_{e,f} \right] \ln{2(2T_eE)^{1/2} \over \hbar\omega_p}, \quad {(\hbar\omega_p)^2 \over T_e} \ll E \ll m_ec^2, \\
L_{si} &=& \ln{2l_sp\over\hbar} - 0.234 - 0.659v^2/c^2, \\
L_{se} &=& L_{si} -{1\over2}\ln{\gamma+3\over2}.
\end{subeqnarray}
$n_{e,f}$ is the free electron density,
$\langle Z^2\rangle=\sum_if_iZ_i^2$ and we use the same ion species
notation as after Eq.\ \ref{eq:zeff} except $Z_i$ is the nuclear (not ionic)
charge. $l_s$ is a screening length, which we take to be the free
electron Debye length. For neutral atoms, it should be replaced by the
atomic radius.  In any event, we impose a minimum of 1 on $L_d$,
$L_{si}$, and $L_{se}$.

We run \zuma\ in 2D cylindrical geometry.  \zuma\ currently operates
with constant (but different) grid spacings $dr$ and $dz$.  The
\hydra\ plasma conditions are interpolated onto a uniform mesh with
$dr=dz=3\ \mu$m using the OVERLINK package
\citep{grandy-overlink-jcp-1999}.  This small spacing is needed to
resolve small features, such as the gold wall and DT layer. The \zuma\
time step is 1 fs, which is chosen to adequately resolve the
dependence of $dE/dt$ on $E$ for small $E$ and high $n_e$.  \zuma\ stops following
electrons when $E<$ 5.11 keV and locally deposits their kinetic energy. In this paper, \zuma\ injects hot electrons from
a distribution that is a product of an energy spectrum $dN/dE$ times a
polar angle spectrum $dN/d\Omega$. For a thermal spectrum with a
``temperature'' $T_h$, we use a relativistic Maxwell-J\"uttner
distribution:
\begin{equation}
  \label{eq:2}
  {dN\over dE} = C\left[ 1+\epsilon/2 \right]^{1/2} \left[ 1+\epsilon \right]
  E^{1/2} e^{-E/T_h}.
\end{equation}
$C$ is a normalization constant, and the two bracketed factors are absent for a
non-relativistic Maxwellian.

\section{Mono-energetic electron propagation through capsule at peak power:
  CAPTEST series}
\label{sec:captest}

\begin{figure}
  \centerline{\includegraphics[width=11cm]{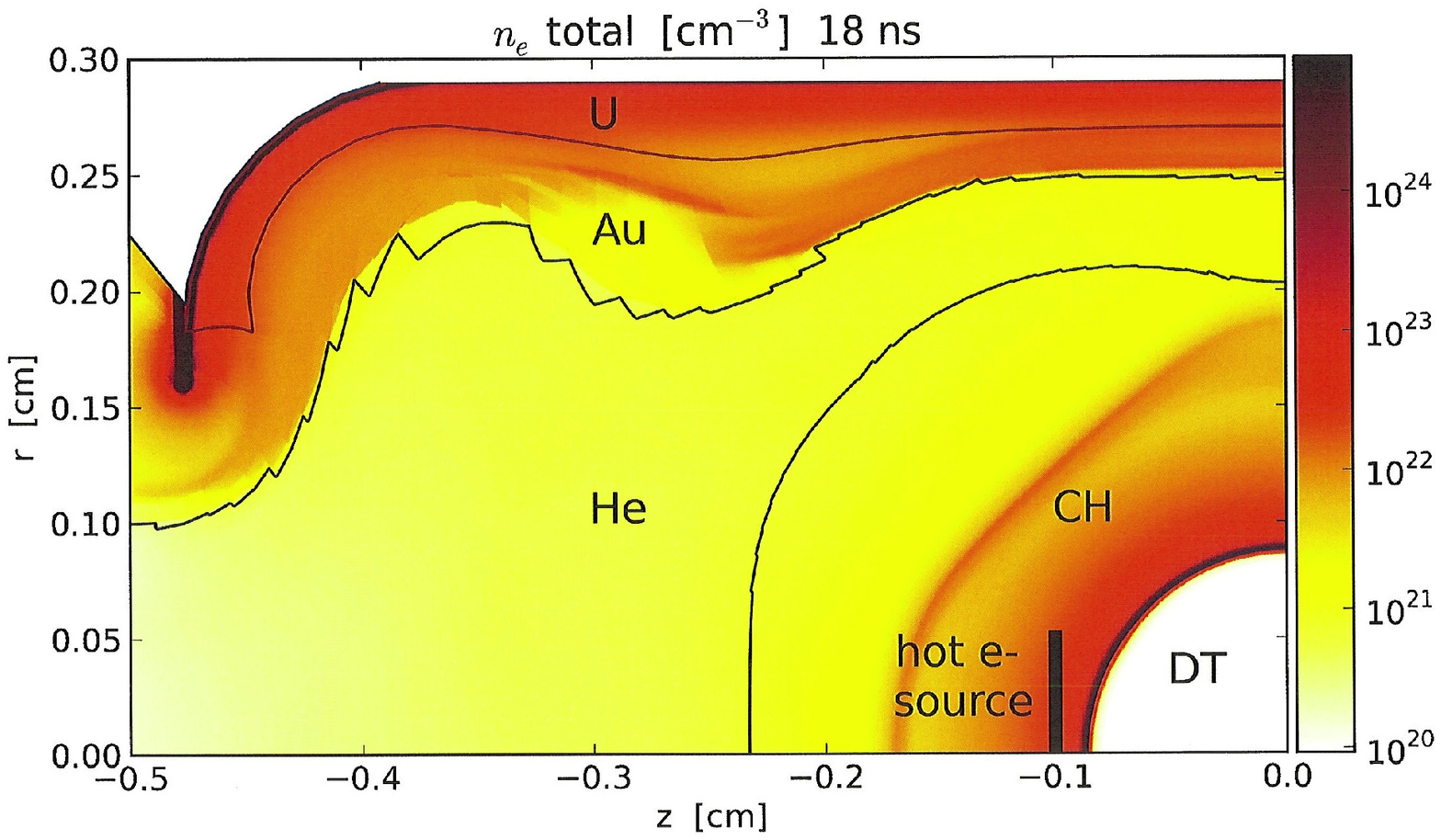}}
  \centerline{\includegraphics[width=11cm]{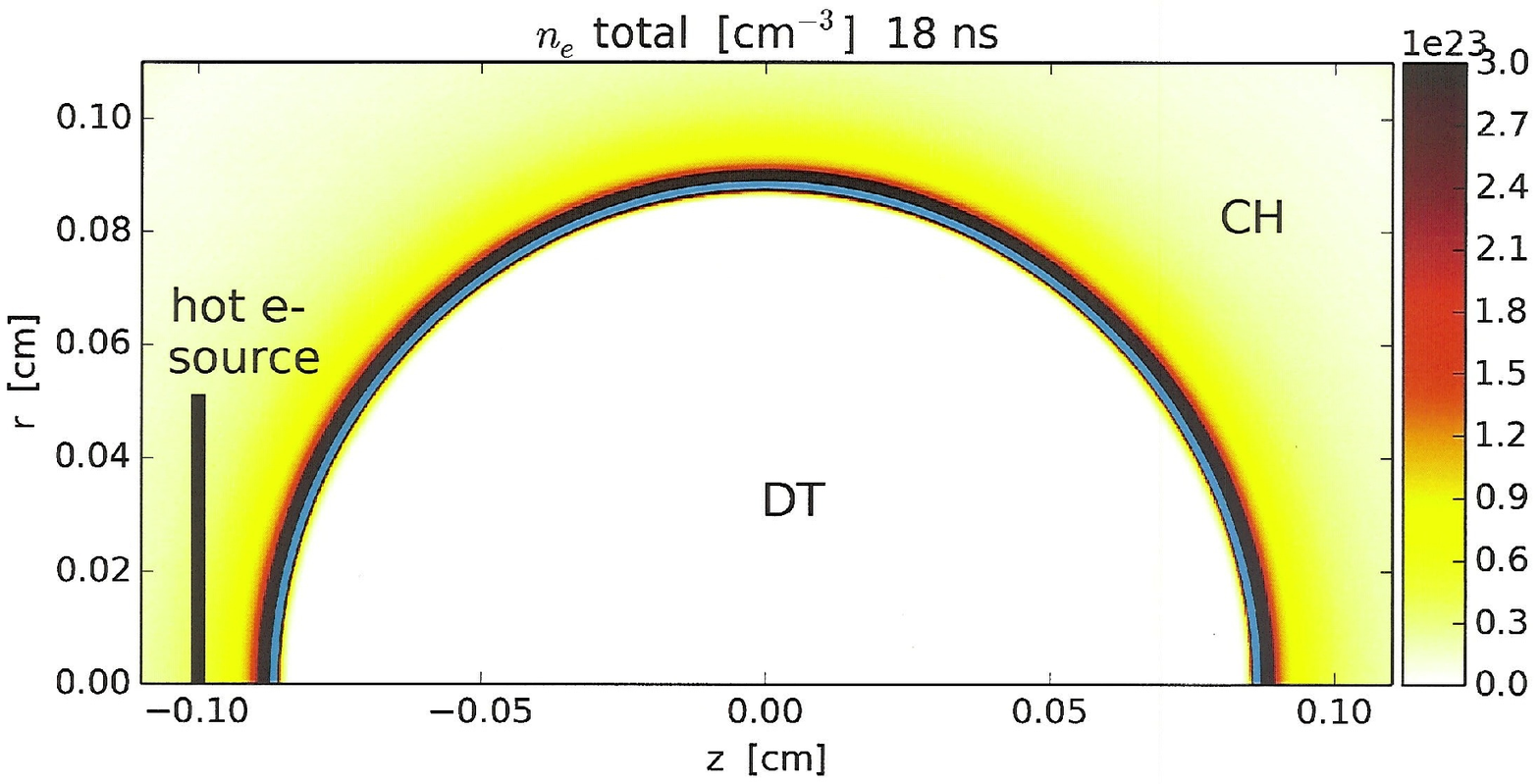}}
  \caption{Electron density (free plus bound) at 18 ns in \hydra\
    simulation of NIF shot N120321 with no MHD (discussed in Sec.\
    \ref{sec:mhdres}), used for CAPTEST series of \zuma\ runs.  Bottom
    panel zooms on capsule. Material regions are separated by solid
    lines and labeled with text as in Fig.\ \ref{fig:18nsmaps}.  The \zuma\ hot electron source is
    indicated.}
\label{fig:18ns-necap}
\end{figure}

This section considers the propagation of electrons directly incident
on the capsule during peak laser power, as a function of electron
energy.  We call this the CAPTEST series of \zuma\ runs, and stress
this source is \textit{not} realistic for LPI-generated hot electrons.
Rather, our purpose is to understand where electrons that reach the
capsule deposit their energy, and which energies pose the greatest
preheat risk. We use plasma conditions from our \hydra\ simulation of
NIF shot N120321 with no MHD at time 18 ns. The same conditions
are used in the SRS-relevant SRSPEAK series discussed below in section
\ref{sec:srspeak}. The time 18 ns is during the rise to peak power
(see Fig.\ \ref{fig:pulse}) and has significant inner-beam SRS.  An
analogous time in shock-timing (``keyhole'') shots has been identified
as possibly having a large hot-electron preheat effect
\citep{robey-visar-pop-2014}.
\begin{figure}
  \centerline{\includegraphics[height=5cm]{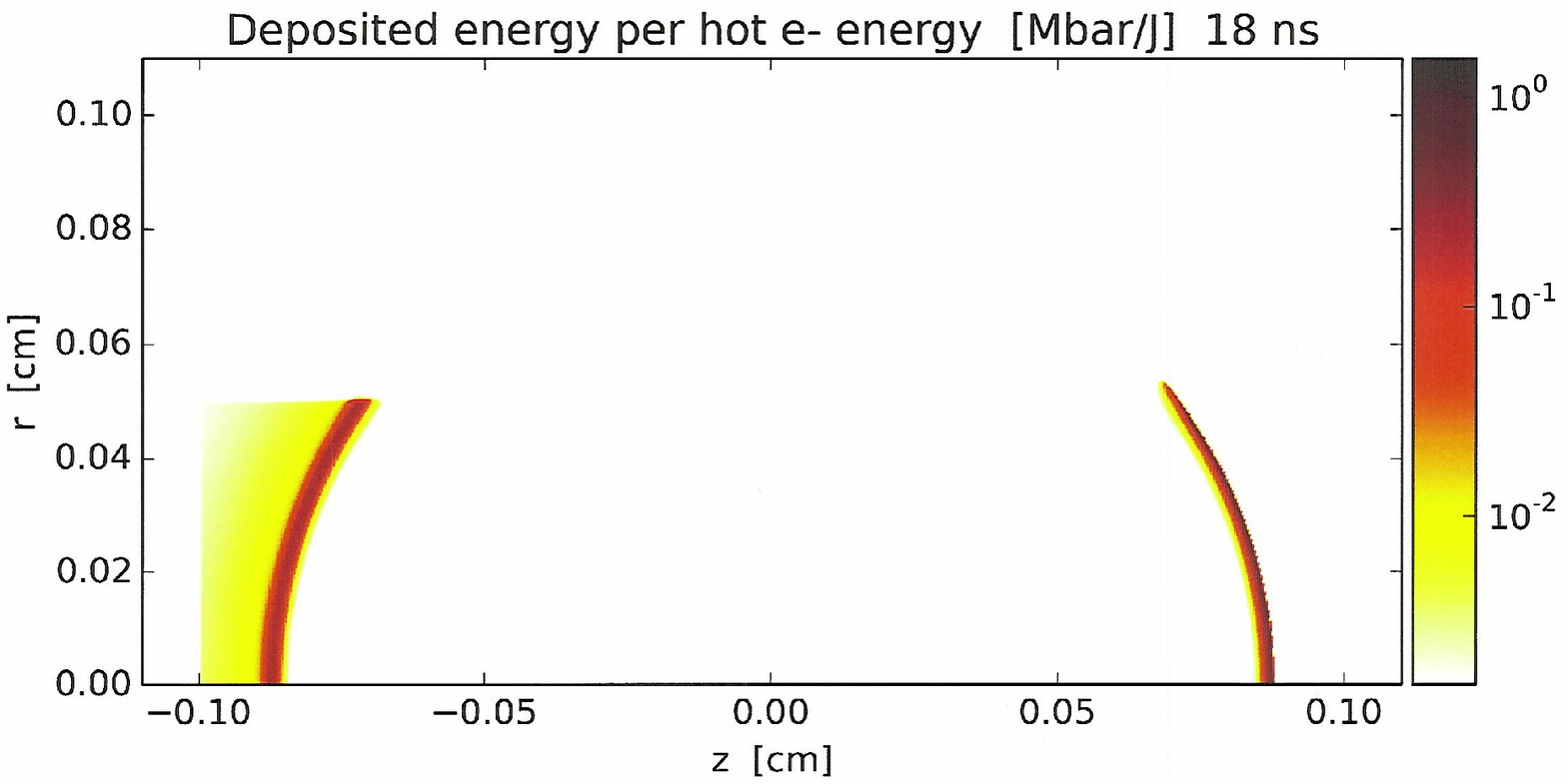}}
  \centerline{\includegraphics[height=5cm]{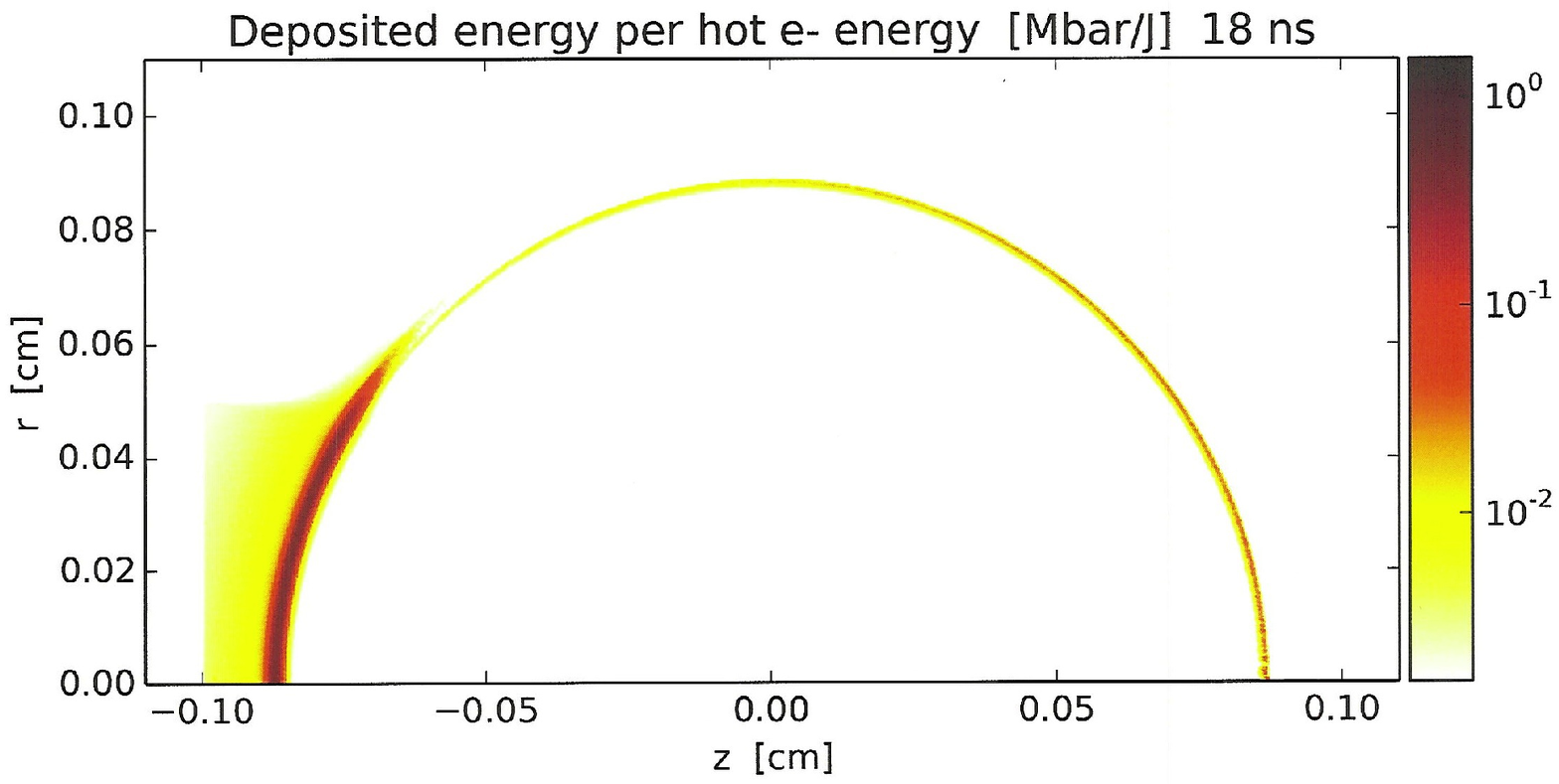}}
  \caption{Energy deposited by 175 keV mono-energetic hot electron source,
    without (top) and with (bottom) angular scattering, for CAPTEST \zuma\ run series.}
\label{fig:eheat-175kv}
\end{figure}
 
Figure \ref{fig:18ns-necap} shows the total (free plus atomically bound) electron
density in the \hydra\ simulation. Mono-energetic hot electrons are injected in a cylinder of
radius 500 $\mu$m at $z=-0.1$ cm, with an initial velocity in the $z$ direction. The hot electrons experience energy loss and
(in some runs) angular scatter, but no forces from $E$ or $B$ fields.  The resulting energy deposited per volume, zoomed on the capsule, is plotted in Fig.\
\ref{fig:eheat-175kv}.  The case with angular scattering shows large spreading
of the hot electrons in the dense CH ablator.  Since the absolute number of hot
electrons introduced is arbitrary, we express the deposition as energy density per
injected hot electron energy.  

\begin{figure}
  \centerline{\includegraphics[angle=90,height=5cm]{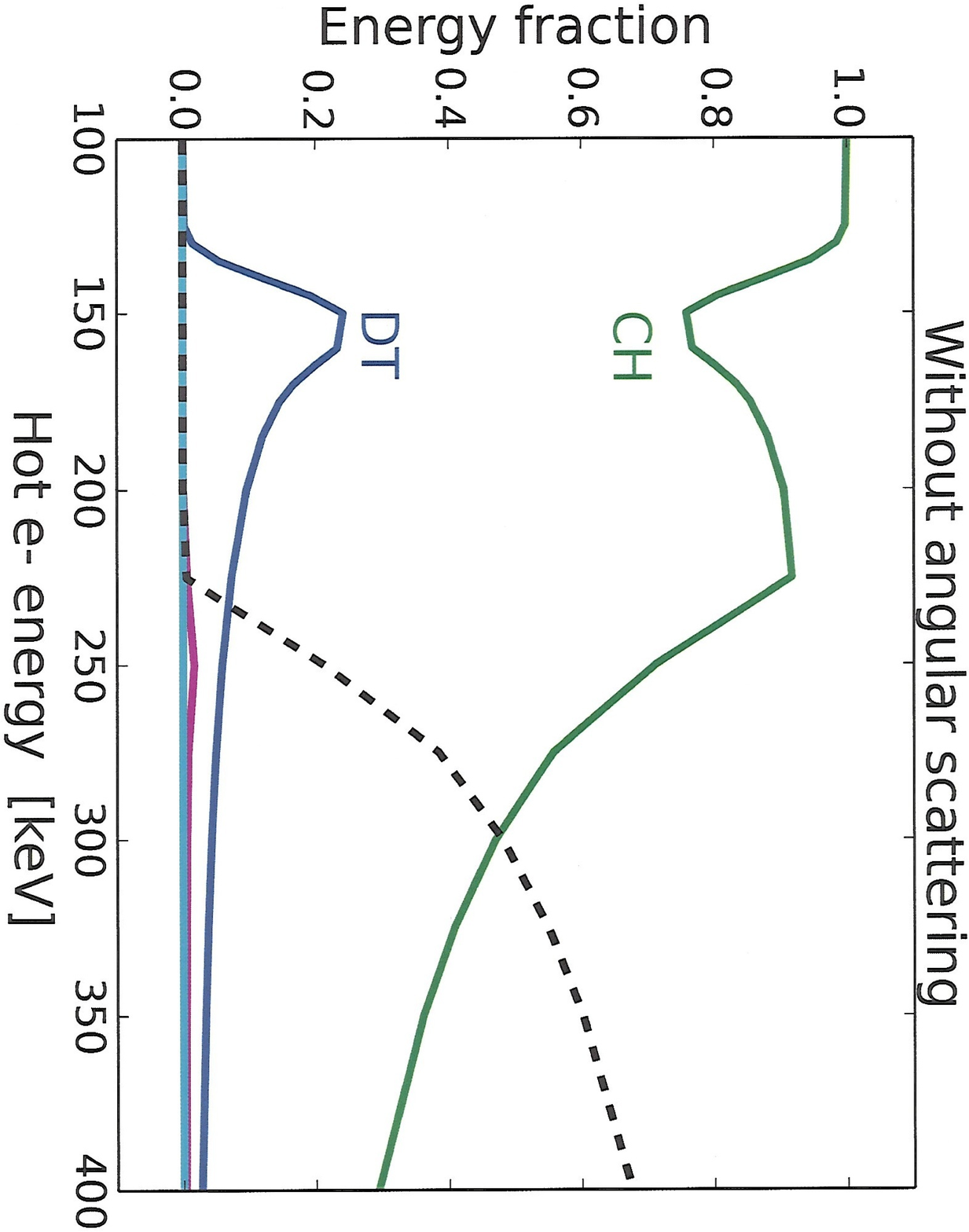} \\
  \includegraphics[height=5cm]{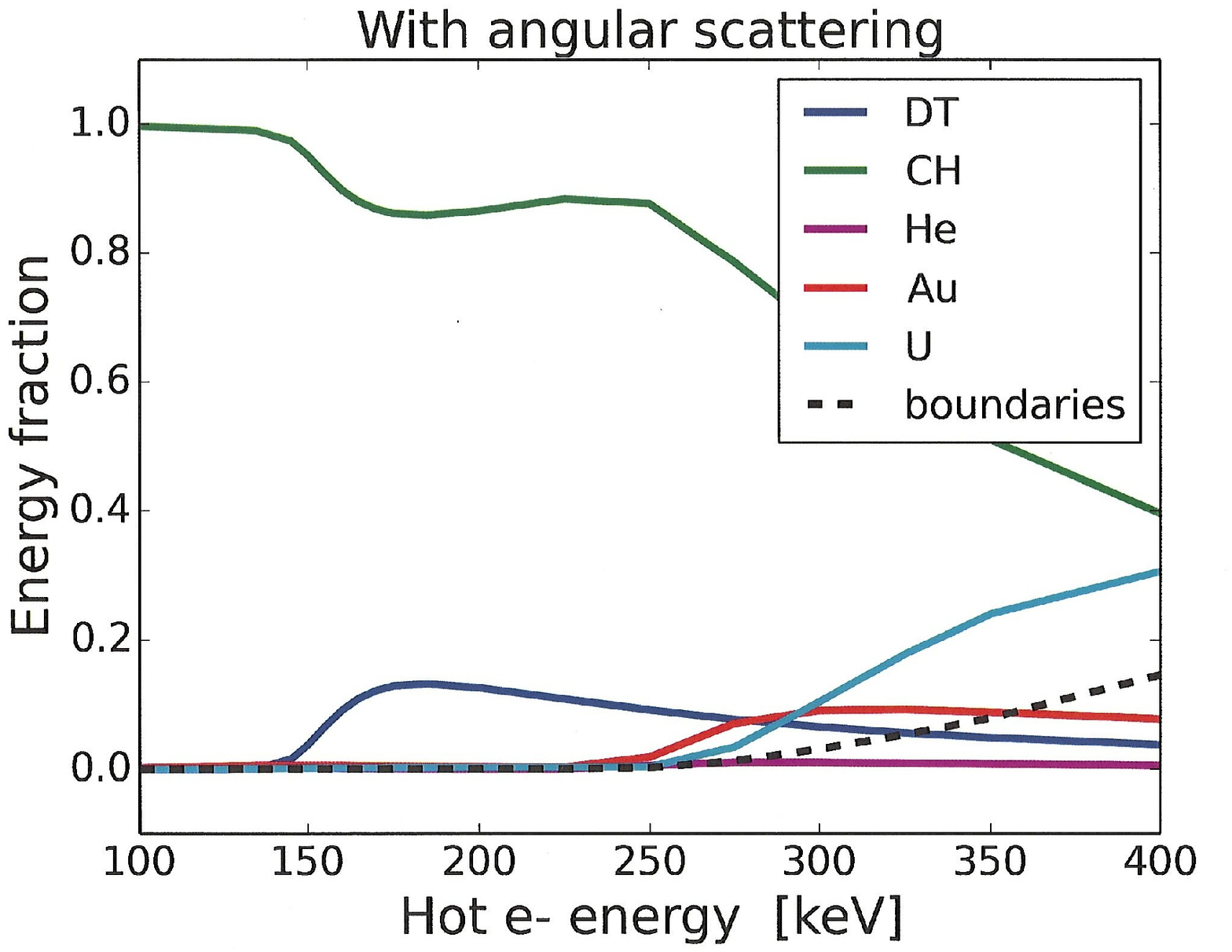}  }
  \caption{Fraction of injected hot electron energy deposited in different
    regions, or escaped to boundaries, for CAPTEST \zuma\ run series.  Plot on (left, right) is (without, with)
  angular scattering.}
\label{fig:mono-frac}
\end{figure}

\begin{figure}
  \centerline{\includegraphics[height=6cm]{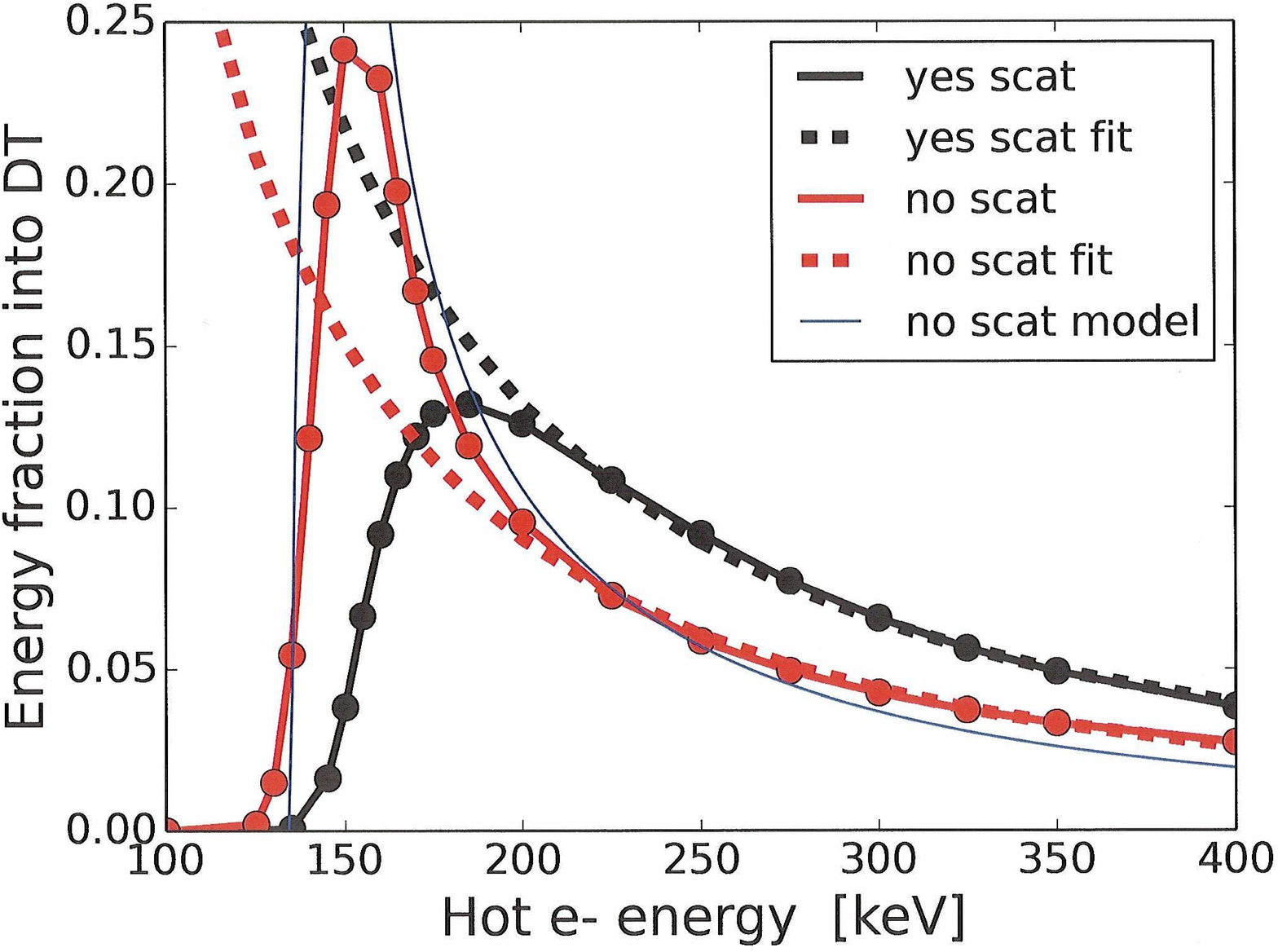}}
  \caption{Solid curves: $\phi_{DT}=$ fraction of injected hot electron energy deposited in DT, with
    (black) and without (red) angular scatter, for CAPTEST \zuma\ run series. Closed circles are \zuma\
    simulation points. Dashed curves: power-law fits to $E\geq200$ keV points.
    Thin blue curve is approximate form explained in text.}
\label{fig:mono-fracDT}
\end{figure}
 
We plot the fraction of injected energy that is
deposited in various regions, or escapes to the boundaries, in Fig.\
\ref{fig:mono-frac}. The deposition in DT (essentially the dense fuel layer,
not the less-dense proto-hotspot) is shown in Fig.\ \ref{fig:mono-fracDT}.  All electrons with
$E<125$ keV stop in the CH ablator.  This sets a minimum
energy hot electrons must have \textit{when they reach the capsule} (at the time
18 ns) to reach the DT layer.  Above this energy, the fraction deposited in DT
increases, until the hot electrons have enough energy to not stop in the
DT.  With no angular scatter (left panel in Fig.\ \ref{fig:mono-frac}), hot
electrons eventually cross the capsule, and exit the problem through the LEH.  Angular scatter lowers and spreads out the peak in
coupling to DT.  It also causes some hot electrons to reach the Au/U hohlraum
wall.  In both cases, the deposition in the He hohlraum fill gas is
negligible.

A simple model illustrates the basic features of Fig.\
\ref{fig:mono-fracDT}, especially for no angular scatter. Imagine a
hot electron starting at position $z_0$ in the CH ablator, with
initial energy $E_0$ and $v_z>0$. We use a 1D slab geometry with CH
from $z_0$ to $z_1$, DT from $z_1$ to $z_2$, and CH for $z>z_2$. We
seek the fraction of initial energy deposited in DT,
$\phi_{DT}=E_{DT}/E_0$, where $E_{DT}$ is the energy deposited in
DT. The hot electron loses energy as it moves to increasing $z$
according to $dE/dz=-f/2E$, where $f$ is a constant, and we include
only the leading-order dependence of stopping power on energy, for
$E\ll m_ec^2$. Integrating from $z_a$ to $z_b$ gives
$E_b^2=E_a^2-f(z_b-z_a)$. An electron with $E_0<E_{01}$ fully stops in
the CH with $z<z_1$, one with $E_0>E_{02}$ crosses the DT, i.e.\ stops
at $z>z_2$, and one with $E_{01}<E_0<E_{02}$ stops in the DT layer,
i.e.\ $z_1<z<z_2$. A straightforward calculation gives
\begin{equation}
  \phi_{DT} = \left\{
    \begin{array}{ll}
      0, & E_0<E_{01} \\[2pt]
      [1-E_{01}^2/E_0^2]^{1/2},   & E_{01}<E_0<E_{02} \\[2pt]
      [1-E_{01}^2/E_0^2]^{1/2} - [1-E_{02}^2/E_0^2]^{1/2},  & E_{02}<E_0.
    \end{array} \right.
\end{equation}
$E_{01}^2=f_{CH}(z_1-z_0)$ and
$E_{02}^2=E_{01}^2+f_{DT}(z_2-z_1)$. For $E_0=E_{01}+\delta E$ with
$\delta E$ small, $\phi_{DT}\approx(2\delta E/E_{01})^{1/2}$, and for
$E_0\gg E_{02}$, $\phi_{DT} \approx f_{DT}(z_2-z_1)/2E_0^2$. The
simple model for $\phi_{DT}$, with $E_{01}=135$ keV and $E_{02}=155$
keV, is plotted as the solid blue curve in Fig.\
\ref{fig:mono-fracDT}. The model is close to the red, no-scattering
result, though the capsule curvature smears the peak compared to the
simple model.  Figure \ref{fig:mono-fracDT} also includes least-square
power-law fits to the $E_0\geq 200$ keV results:
$\phi_{DT}=(E_0/52.3\ \mathrm{keV})^{-1.79}$ without angular scatter,
and $\phi_{DT}=(E_0/62.7\ \mathrm{keV})^{-1.75}$ with scatter. These
are both close to the $E_0^{-2}$ scaling of our simple model for
$E_0\gg E_{02}$.

\begin{figure}
  \centerline{\includegraphics[height=6cm]{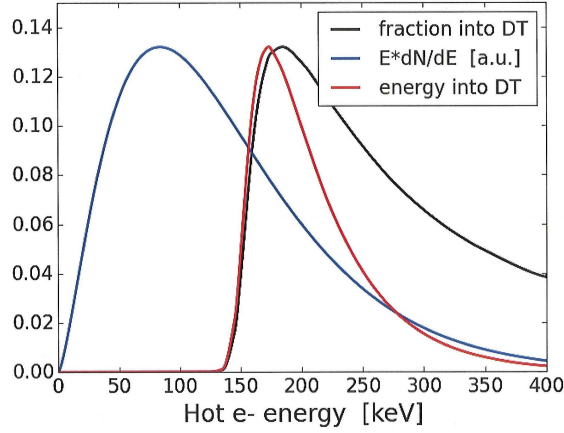}}
  \caption{Coupling to DT of a Maxwell-J\"uttner hot electron source. Blue:
    $E*dN/dE$ for J\"uttner source with $T_h=50$ keV.  Black: $\phi_{DT}=$
    energy fraction deposited in DT, with angular scatter (solid black curve from Fig.\
    \ref{fig:mono-fracDT}). Red: product of black and blue curves.}
\label{fig:mono-Thot1}
\end{figure}

\begin{figure}
  \centerline{\includegraphics[height=6cm]{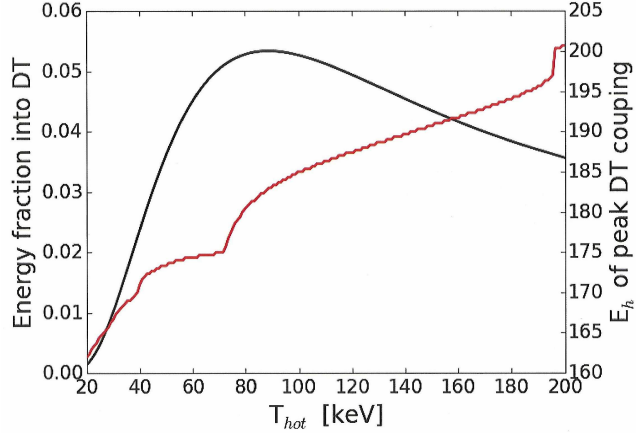}}
  \caption{Black (left $y$-axis): injected hot electron energy fraction coupled to DT, for
    Maxwell-J\"uttner source with temperature $T_{hot}$.  Red (right $y$-axis): hot
    electron energy with peak coupling to DT (i.e., peak of red curve from Fig.\
    \ref{fig:mono-Thot1}).}
\label{fig:mono-Thot-funcs}
\end{figure}

We apply our mono-energetic results to a thermal spectrum in Figs.\
\ref{fig:mono-Thot1} and \ref{fig:mono-Thot-funcs}, and find DT
preheat comes mainly from hot electrons with energies $\gtrsim 160$
keV, for $T_h>20$ keV. Figure \ref{fig:mono-Thot1} shows the coupling
to DT of a thermal, Maxwell-J\"uttner spectrum with $T_h=$ 50 keV. The
black curve is the DT coupling fraction from Fig.\
\ref{fig:mono-fracDT} (black curve there too), and the blue curve is
the thermal energy spectrum $E*dN/dE$ for $T_h=50$ keV. The red curve
is their product, namely the energy coupled to DT by electrons of a
given energy, in a thermal spectrum. The red curve exhibits behavior
akin to the ``Gamow peak'' in fusion reactions, with a location
determined essentially by the steeper black curve. Figure
\ref{fig:mono-Thot-funcs} shows the overall $\phi_{DT}$ integrated
over the thermal spectrum vs.\ $T_h$. This peaks slightly above 5\%
near $T_h=90$ keV. The red curve is, as a function of $T_h$, the hot
electron energy of maximum $E_{DT}$, i.e.\ the energy of the peak in
the red curve in Fig.\ \ref{fig:mono-Thot1}. This increases slowly with
$T_h$, and is at $>160$ keV for all $T_h$ of interest. It is thus
important to correctly model these hot electrons to calculate DT
preheat, even for $T_h\ll 160$ keV.

\section{Hot electron propagation in early-time picket: PICKET series}
\label{sec:picket}

This section studies hot electron dynamics during the initial laser
``picket,'' and the effects of an axial $B$ field.  The principal way
hot electrons are produced during the picket is LPI in the LEH.  This
can be two-plasmon decay (TPD) for $n_e \approx n_{cr}/4$, Raman
scattering, or a multi-beam variant of it
\citep{michel-collsrs-prl-2015}. NIF experiments have shown the picket
hot electrons can be reduced by shaping the picket pulse, for instance
by turning the inner beams on before the outers to blown down the
window at low power. Experiments at the Omega laser studied hot
electrons from TPD during window burn-down
\citep{regan-hote-pop-2010}.

The DT fuel is particularly sensitive to hot electrons produced during
the early time picket pulse: $\Delta$entropy =$E_{DT}$/temperature, so
a small $E_{DT}$ added when the fuel is cold produces a large entropy
increase.  In addition, melting the cryogenic DT layer before the
first shock arrives causes the inside surface to expand, which can
degrade the ability to shock-time \citep{thomas-privcom-hots-2015}.
For indirect-drive ignition designs, this occurs for $E_{DT}\sim0.1$
J.  NIF ignition-relevant hohlraum experiments show total hot-electron
energies $E_h\sim1$ J with $T_h\sim 80$ keV.  Calculations typically
show $E_{DT}/E_h\sim(2-5)\times10^{-3}$, giving preheat
$E_{DT}\sim(2-5)\times10^{-3}$ J well below melt.

\begin{figure}
  \centerline{\includegraphics[height=7cm]{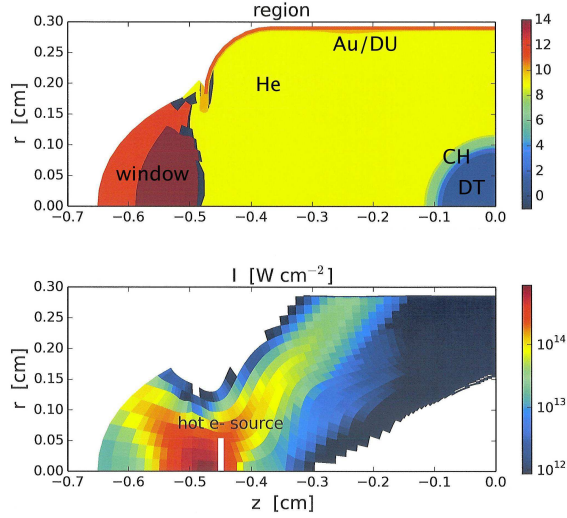}}
  \caption{Material region (top) and laser intensity summed over all beams (bottom) at 1 ns from \hydra\ simulation with no MHD of NIF shot N120321, used for PICKET series of \zuma\ runs. Hot electron source is indicated as white box, and is placed only in the $z<0$ half of the two-sided ($z<0$ and $z>0$) \zuma\ domain.}
\label{fig:pick-maps}
\end{figure}

 We use \hydra\ plasma conditions at 1 ns,
shortly after the outer-beam power has peaked, for \zuma\ calculations. We call
this the PICKET run series.  Figure \ref{fig:pick-maps} shows the material
regions and laser intensity. We source the hot electrons in a 500 $\mu$m radius
circle at the left-side LEH ($z=-0.45$ cm), which is roughly the extent of high
laser intensity.  Since TPD does not generally produce collimated hot electrons,
we use an isotropic source with velocity-space $dN/d\Omega$ constant ($\Omega$
is solid angle in velocity) for polar angles between 0 and 90$^\circ$, and zero otherwise
(i.e., uniform in the forward-going half-space). The energy spectrum is a
Maxwell-J\"uttner with $T_h=80$ keV, which is consistent with hard x-ray data on NIF (discussed below).

\begin{figure}
  \centerline{\includegraphics[height=6cm]{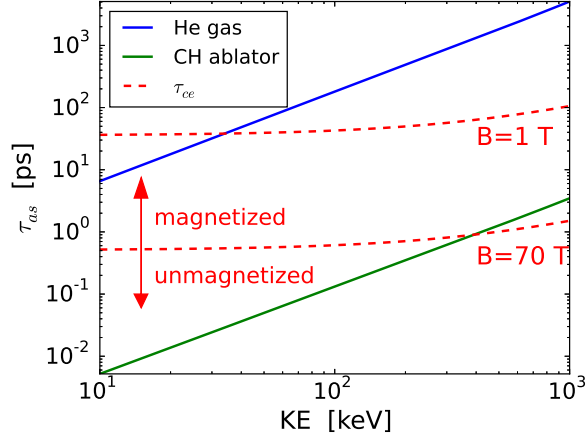}}
  \caption{Time for $90^\circ$ angular scatter $\tau_{as}$ for: blue curve: 0.96 \mgcc\ of He at $T_e$=1 keV (typical for hohlraum gas fill in NIF shot N120321), green curve: 1 \gcc\ of $C_1H_1$ at 200 eV (typical for the ablator). Red dashed curves are cyclotron period $\tau_{ce}$ for $B_z=1$ and 70 T.  Most hot electrons are (strongly, weakly) magnetized in the (He gas, CH ablator).}
\label{fig:pick-tauas}
\end{figure}

We expect a 70 T axial field to strongly magnetize the hot electrons in
the low-density hohlraum gas fill, and guide them to the capsule.  Recall that we
inject a divergent hot electron source, so the question is whether the field
confines them in space.  It will not collimate them, i.e.\ reduce their
velocity-space divergence. The electron Larmor radius $r_{Le}\equiv p_\perp/eB$, which for
$E=100$ keV and $B=70$ T is $r_{Le}=16.0\ \mu$m$\cdot\sin\alpha$ ($\alpha$ is the angle
between $\vec B$ and $\vec p$). This is much less than the relevant plasma scale
lengths. Also, the cyclotron period $\tau_{ce}=2\pi\gamma m_e/eB$ is 0.510
ps$\cdot\gamma$ for $B=70$ T, which is much shorter than the
propagation time through the hohlraum. Figure \ref{fig:pick-tauas} plots $\tau_{ce}$ and the time for 90$^\circ$ root-mean-square angular scatter, $\tau_{as}$: $\left\langle \theta^2 \right\rangle=(90^\circ)^2$ for $dt=\tau_{as}$ in Eq.\ \ref{eq:dta2dt}. We consider two fully-ionized cases: one
representative of the hohlraum fill: 0.96 \mgcc\ of He at $T_e$=1 keV, and one
of the ablator: 1\ \gcc\ of $C_1H_1$ at 200 eV. The plot shows all hot electrons
are magnetized in the He, while those with $E>$ 300 keV in the CH are.
Even if $\tau_{ce}$ exceeds $\tau_{as}$, $r_{Le}$ is much
smaller than typical capsule dimensions $\sim$ 100's $\mu$m.

\begin{figure}
  \centerline{\includegraphics[height=7cm]{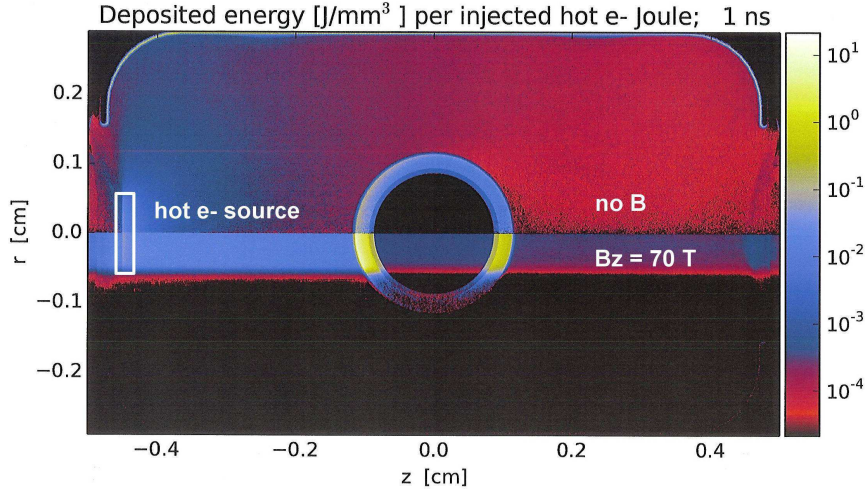}}
  \caption{Hot electron energy deposition for PICKET series of \zuma\ runs with no magnetic field (top) and a uniform $B_z=70$ T field (bottom).}
\label{fig:pick-edep}
\end{figure}

The energy deposition is shown in Fig.\ \ref{fig:pick-edep} for the
ZUMA runs with no $B$ field (top half), and with a uniform $B_z=$70 T
field (bottom half).  Table \ref{tab:edep} lists the fraction of
injected hot electron energy deposited in different regions.  With no
$B$ field, the hot electrons propagate essentially freely in the He
gas fill. They mostly deposit in the hohlraum wall, and a small
fraction deposits in the ablator.  This is expected based on the solid
angle subtended by these regions.  With a uniform $B_z=70$ T, the hot
electrons are strongly magnetized in the He gas and guided to the
capsule.  They mostly deposit in the ablator, out to a radius
comparable to that of the source.  This asymmetric preheat, occurring
mostly in the poles, may drive capsule asymmetries.  The energy deposited in the DT layer is $\sim12$x higher
with the 70 T field.  Whether this is a preheat concern depends on the
spectrum and total energy of hot electrons produced.

The lack of deposition in high $Z$ with the field means the same hot
electron source produces many fewer hard x-rays.  This is a
diagnostics concern, since hard x-rays are generally used to deduce
hot electrons on NIF. One such principal diagnostic is the FFLEX hard
x-ray ($>$ 10 keV) detector \citep{dewald-fflex-rsi-2010,
  hohenberger-fflex-rsi-2014}, with 10 channels filtered for different
energy ranges. Energetic electrons lose energy by collisions with
background electrons and by bremsstrahlung radiation.  Radiation loss
$\sim EZ \times$ collisional loss, with the two equal in gold for
$E=10$ MeV.  Only electrons that deposit energy in high-$Z$ material,
such as the hohlraum wall, produce enough hard x-rays for FFLEX to
detect. Hot electrons striking the capsule during the picket have been
measured on ``re-emit'' experiments, where the capsule is replaced by
a high-$Z$ (e.g.\ bismuth) ball.

\begin{table}
  \begin{center}
\def~{\hphantom{0}}
  \begin{tabular}{l|c|c|c|c}
      Region      & PICKET,    & PICKET,           & SRSPEAK 1, & SRSPEAK 1,   \\
                  & no B       &  $B_z=70$ T       & no MHD     & $B_{z0}=70$ T \\ [3pt]
       DT gas     & 6.56E-5    & 1.06E-3 (16x)     & 4.32E-6    & 6.26E-5 (14x)      \\
       DT layer   & 2.20E-3    & 0.0261  (12x)     & 3.58E-4    & 2.89E-3 (8.1x)      \\
       CH ablator & 0.0749     & 0.696   (9.3x)    & 0.406      & 0.804   (2.0x)    \\
       He gas     & 0.0566     & 0.0646  (1.1x)    & 0.223      & 0.117   (0.52x)   \\
       Au         & 0.366      & 4.14E-4 (1.1E-3x) & 0.250      & 1.01E-4 (4.0E-4x) \\
       DU         & 0.428      & 4.02E-4 (9.4E-4x) & 0.0990     & 1.61E-5 (1.6E-4x) \\
       total      & 0.927      & 0.789   (0.85x)   & 0.979      & 0.925   (0.94x)  \\
\hline \\
      Region      & SRSPEAK 2, & SRSPEAK 2,        & SRSPEAK 3, & SRSPEAK 3,     \\
                  & no B       & $B_{z0}=70$ T     & no B       & $B_{z0}=70$ T   \\
       DT gas     & 1.75E-6    & 8.95E-9 (5.1E-3x) & 1.44E-6    & 5.96E-6 (4.1x) \\
       DT layer   & 1.37E-4    & 3.44E-6 (0.025x)  & 1.19E-4    & 1.26E-3 (11x)  \\
       CH ablator & 0.272      & 0.105   (0.39x)   & 0.327      & 0.576 (1.8x)   \\
       He gas     & 0.229      & 0.499   (2.2x)    & 0.182      & 0.248 (1.4x)   \\
       Au         & 0.335      & 0.220   (0.66x)   & 0.328      & 0.101 (0.31x)  \\
       DU         & 0.133      & 0.0421  (0.032x)  & 0.131      & 5.56E-3 (0.042x) \\
       total      & 0.969      & 0.866   (0.89x)   & 0.968      & 0.932 (0.96x)
  \end{tabular}
  \caption{Fraction of injected hot electron energy deposited in different regions, for PICKET (1 ns) and SRSPEAK (18 ns) series of \zuma\ runs. (x) is ratio of with $B$/MHD to without. Fractions do not sum to unity because some hot electrons escape from problem boundaries.}
  \label{tab:edep}
  \end{center}
\end{table}

\section{Hot electron propagation during peak power: SRSPEAK series}
\label{sec:srspeak}

We now consider \zuma\ simulations of propagation of a realistic hot
electron source produced by Raman scattering on the inner beams during
peak power. We use the same plasma conditions at 18 ns that were used
in the CAPTEST series (i.e., a simulation of NIF shot N120321 with the
full, incident laser power on each cone), along with conditions from a
\hydra\ run with an initial $B_{z0}=70$ T axial field and the MHD
package active. The hot electron source has a Maxwell-J\"uttner energy
spectrum with $T_h=$ 30 keV. This temperature is gotten from FFLEX
data at 18 ns (rise to peak power) on NIF shot N130517, which is
analogous to N120321 \citep{robey-visar-pop-2014}.  Once peak power is
reached, $T_h=$ 18 keV is consistent with FFLEX data. The injected
angle spectrum is
$dN/d\Omega = \exp[-((\theta-27^\circ)/10^\circ)^4]$, which is
directed along roughly the bisector of the two NIF inner beams at
$\theta=23.5^\circ$ and 30$^\circ$.  

\begin{figure}
  \centerline{\includegraphics[height=6cm]{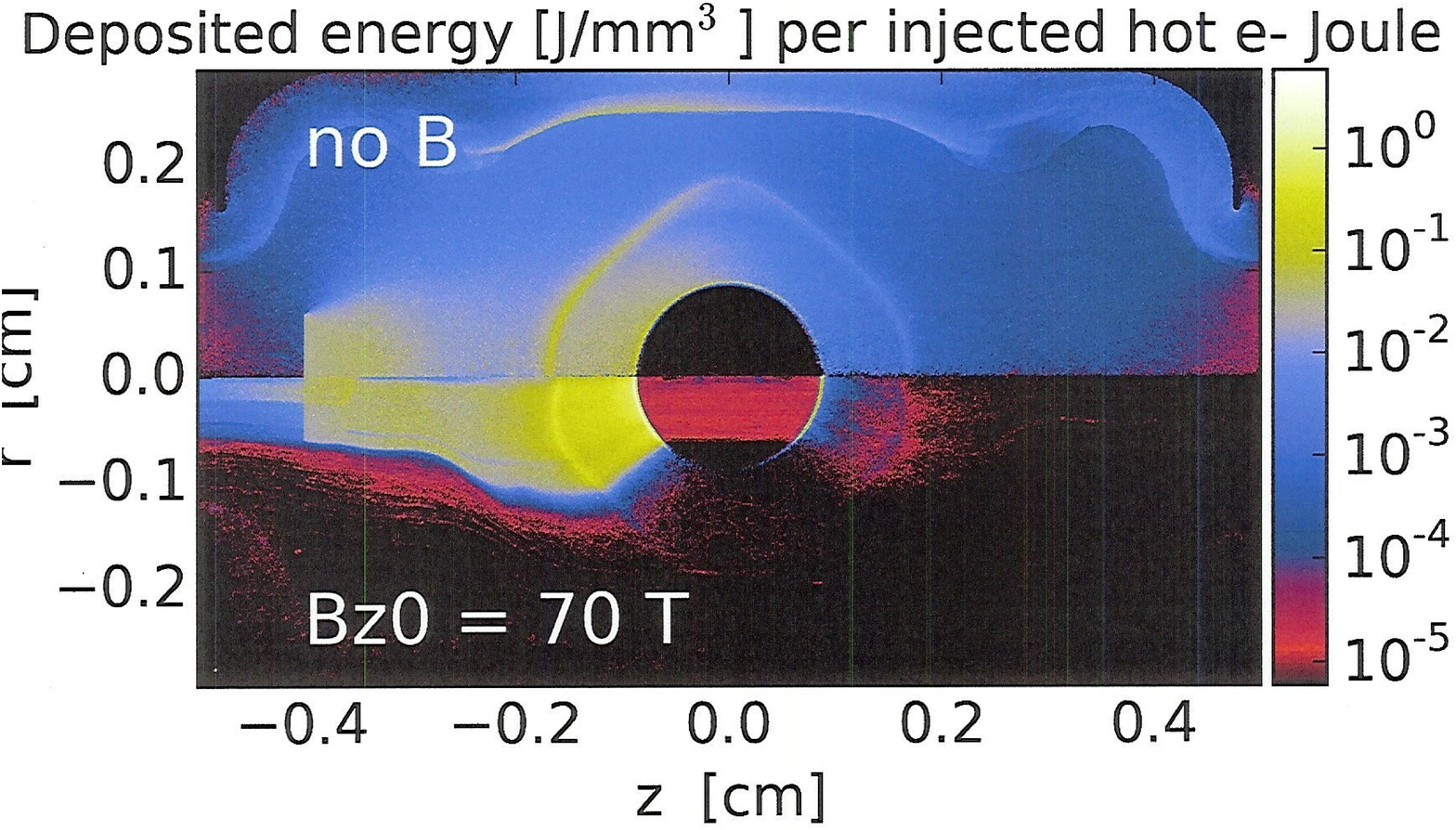}}
  \centerline{\includegraphics[height=6cm]{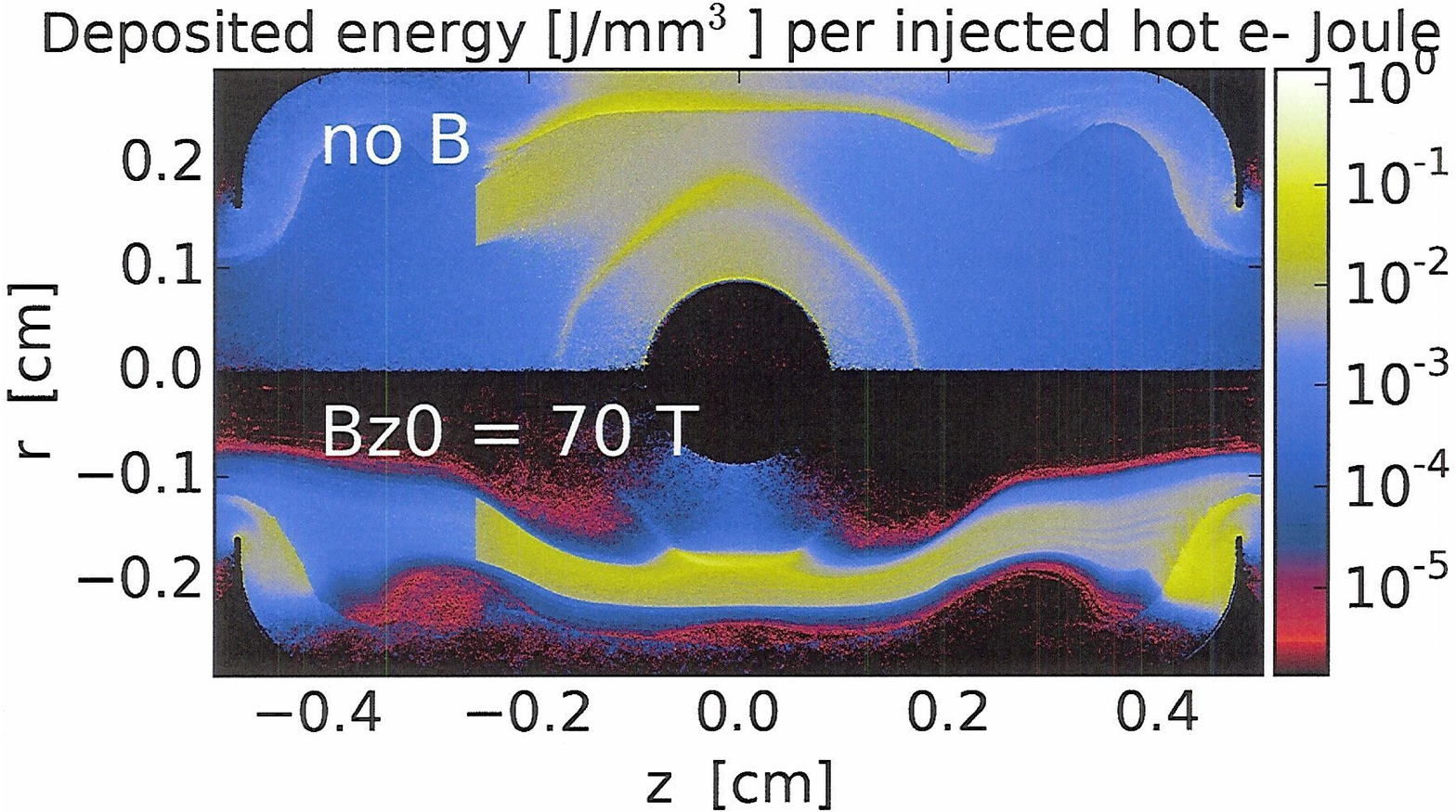}}
  \centerline{\includegraphics[height=6cm]{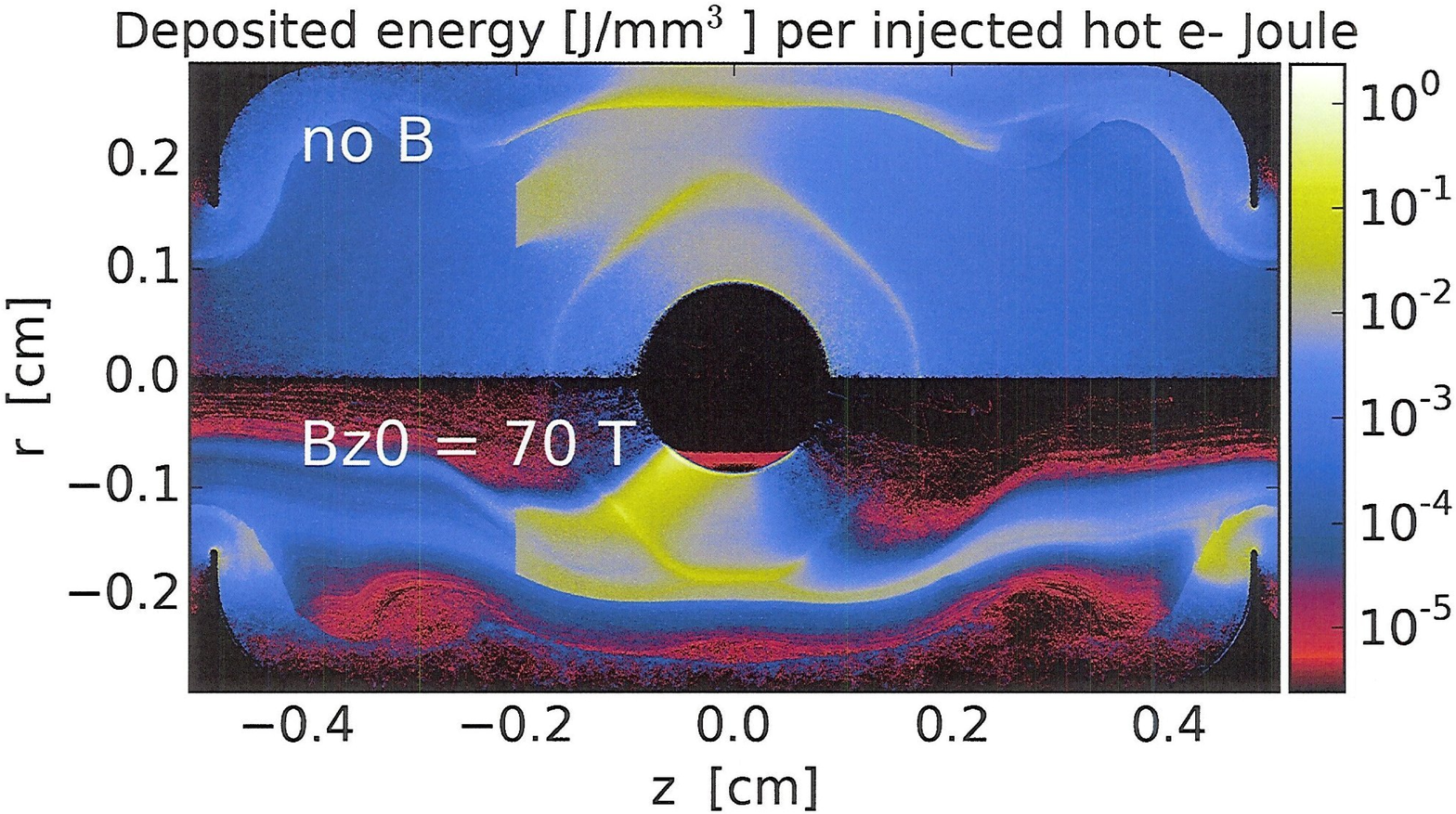}}
  \caption{Hot electron energy deposition for SRSPEAK series of \zuma\ runs without MHD (top $r>0$) and with MHD (bottom $r<0$). Top, middle, and bottoms plots are for source locations 1, 2, 3 indicated in Fig.\ \ref{fig:18nsB}.}
\label{fig:srs-edep}
\end{figure}

We find strong sensitivity to what field lines hot electrons start on
-- namely, whether or not the field lines connect to the capsule. Hot
electrons are injected in the three locations indicated as sources 1,
2, and 3 in Fig.\ \ref{fig:18nsB}: from $r=0$ to 0.06 cm at $z=-0.4$
cm, and from $r=$ 0.12 to 0.18 cm, at $z=-0.2$ and -0.25 cm.  The
energy deposition vs.\ space is plotted in Fig.\ \ref{fig:srs-edep},
and the total into various materials is given in Table
\ref{tab:edep}. With no MHD, the fraction of hot electron energy
deposited to DT varies from $(1.2-3.6)\times 10^{-4}$ over the three
sources. The field strongly magnetizes the hot electrons in the He
fill gas for all three sources, as in the PICKET series of Section
\ref{sec:picket}. Also like the PICKET series electrons from source 1,
in the LEH, are guided to the capsule. The deposition in DT, CH, and
He is greatly increased compared to the no-MHD case. For source 2,
located deeper in the hohlraum and off axis, electrons are injected on
field lines that do not connect to the capsule. The resulting
deposition in He gas is significantly increased compared to the no-MHD
case, while that into the CH ablator and especially the DT layer are
reduced. The situation reverses for source 3, which is slightly closer
to the capsule in $z$ than source 2.  Some electrons now start on
field lines that connect to the capsule, which results in much higher
DT deposition.  It is not presently known where in the hohlraum SRS
hot electrons are produced, so we cannot say whether the field
increases or decreases DT deposition.  As with the PICKET series, the
fraction of hot electron energy deposited in high-$Z$ material is
lower with the field, especially for source 1. Hard x-ray diagnostics
may thus not be reliable indicators of hot electron preheat.

\section{Conclusions}
\label{sec:conc}

This paper gave results of \hydra\ rad-hydro simulations with no MHD
effects, and with MHD and a 70 T initial axial $B$ field.  The field
is essentially frozen-in to the highly conductive plasma, and gets
advected with the radial expansion of the capsule and wall. This
results in field lines that roughly follow contours of ablated
material. The magnetic pressure is much less than material
pressure. The principal hydro effect of the field is reduced electron
heat conduction perpendicular to it. This gives a hotter hohlraum
fill, especially in gold, and a wider channel between the capsule and
equator wall. Less inner-beam absorption occurs before they reach the
wall, which increases the equatorial x-ray drive. Inner-beam Raman
scattering may be reduced by the hotter fill, in addition to the lower
power needed to achieve a round implosion.

We also presented hot electron propagation studies with \zuma, using
plasma conditions from \hydra. Mono-energetic test cases with plasma
conditions from early peak power (18 ns) show a minimum hot electron
energy of 125 keV incident on the capsule is required to reach the DT
layer.  The energy coupled to the layer maximizes at 13\% for 185 keV
electrons, and drops with energy above that. Using plasma conditions
during the early-time picket (1 ns) with no field, we find a small
fraction $(2\times 10^{-3})$ of hot electron energy from a
two-plasmon-decay relevant source couples to DT. With a uniform 70 T
axial $B$ field, the hot electrons are magnetized in the He fill gas,
guided to the capsule, and the DT coupling increases by 12x. This may
not be a preheat concern, since picket pulse shaping has been shown on
NIF to significantly decrease the hot electron source.  \zuma\
simulations using plasma conditions at 18 ns, with a source motivated
by inner-beam SRS, show an imposed field can greatly increase or
decrease hot-electron coupling to DT. This depends on whether
electrons are produced on field lines that connect to the capsule.

Imposed magnetic fields may enhance hohlraum performance by improving
inner-beam propagation and reducing Raman scattering during peak laser
power. This is in addition to the primary benefit of reducing electron-heat and
alpha-particle loss from the hotspot. One concern is possible increase
in DT fuel preheat due to the field guiding hot electrons to the
capsule. Work is underway on a pulsed-power field generator for NIF,
and we look forward to hohlraum experiments in the next few years.

We gratefully acknowledge fruitful conversations with H.\ F.\ Robey,
J.\ D.\ Salmonson, C.\ A.\ Thomas, J.\ Hammer, and D.\ E.\
Hinkel. This work was performed under the auspices of the
U.S. Department of Energy by Lawrence Livermore National Laboratory
under Contract DE-AC52-07NA27344.  Partly supported by LLNL LDRD
project 14-ERD-028.

\bibliographystyle{jpp}

\bibliography{hote14}

\end{document}